\documentclass[fleqn,usenatbib]{mnras}
\usepackage{newtxtext,newtxmath}
\usepackage[T1]{fontenc}
\usepackage{graphicx} 
\usepackage{amsmath}
\usepackage{xcolor}
\usepackage[capitalise]{cleveref}
\usepackage{booktabs}
\usepackage{makecell}
\usepackage{hyperref}

\crefname{equation}{equation}{equations}
\Crefname{equation}{Equation}{Equations}
\DeclareRobustCommand{\VAN}[3]{#2}
\let\VANthebibliography\thebibliography
\def\thebibliography{\DeclareRobustCommand{\VAN}[3]{##3}\VANthebibliography}

\newcommand{\n}{\mathbf{n}}
\newcommand{\T}{\mathbf{T}}
\newcommand{\A}{\mathbf{A}}
\renewcommand{\a}{\mathbf{a}}
\renewcommand{\d}{\mathbf{d}}

\title[Low-$\ell$ 21-cm mapmaking]{Low multipole mapmaking for global 21-cm experiments}
\author[Y. D. Ignatov et al.]{
Yordan D. Ignatov$^{1}$\thanks{E-mail: yi17@ic.ac.uk (YDI)}, Jonathan R. Pritchard$^{1,2}$
\\
$^{1}$Department of Physics, Blackett Laboratory, Imperial College London, London SW7 2AZ, UK\\
$^{2}$Max-Planck-Institut für Radioastronomie, Auf dem Hügel 69, D-53121 Bonn, Germany
}

\date{Accepted XXX. Received YYY; in original form ZZZ}
\pubyear{2024}

\begin{document}
\label{firstpage}
\pagerange{\pageref{firstpage}--\pageref{lastpage}}
\maketitle

\begin{abstract}
The 21-cm global signal is obscured by very bright galactic and extra galactic foreground emissions. Typical single-spectrum fit (SSF) based methods for foreground/signal separation can result in biased estimates of the cosmological signal due to the presence of spectral oscillations induced by the interaction between chromatic beams and the spatial shape of the foregrounds. Modelling this interaction requires some amount of assumed foreground information. We present a mapmaking-based approach which describes the beam-weighted observation of the sky by multiple globally-distributed antenna experiments as an observation equation. This equation is inverted in order to estimate the low-order sky modes ($\ell\lesssim10$). The resulting chromaticity-free sky monopole is then fit with a smooth foreground function and a 21-cm model. Given the insensitivity of global 21-cm experiments to small angular scales, we rely on the mean and covariance of higher-order foreground modes being known. We show that this mapmaking-based method is capable of inferring the cosmological signal in cases where a SSF with a simple beam-factor based chromaticity correction fails, even when the foreground model used in the mapmaking method features uncertainty at the 10\% level.
\end{abstract}


\section{Introduction}
The 21-cm signal, formed in the interaction between the Cosmic Microwave Background (CMB) and neutral hydrogen clouds in the Dark Ages and Epoch of Reionization (EoR), stands as one of the most promising direct probes of the Universe in those time periods. The detection of the sky-averaged 21-cm global signal from the Cosmic Dawn would provide valuable constraints on the timings and properties of galactic formation and the formation and death of the first stellar objects. The detection of the 21-cm signal from the cosmic Dark Ages could provide additional cosmological constraints, helping us understand and potentially resolve current cosmological tensions \citep[see e.g.][for a review]{2012:PritchardLoeb}.

Experimental efforts to detect the 21-cm global signal have focused almost exclusively on the absorption trough in the era of Cosmic Dawn, as the smaller absorption trough formed in the Dark Ages is obscured by much brighter foregrounds. The detection of the absorption trough of the global signal formed in the era of first stars poses significant challenges due to the large galactic and extra-galactic foreground contamination, 1000s to 10000s of times brighter than the signal itself in some bins. Significant experimental effort has been made to detect this signal by a wide variety of current and upcoming experiments, such as EDGES \citep{2018:BowmanRogersMonsalve}, SARAS \citep{2021:NambissanT.SubrahmanyanSomashekar}, PRIZM \citep{2019:PhilipAbdurashidovaChiang}, LEDA \citep{2018:PriceGreenhillFialkov}, REACH \citep{2022:deLeraAcedodeVilliersRazavi-Ghods} and MIST \cite{2023:MonsalveByeSievers}. 

Most notably, the EDGES collaboration published a claimed detection of the Cosmic Dawn absorption signal \citep{2018:BowmanRogersMonsalve}, the magnitude of which was more than twice that predicted by the largest standard-cosmological models \citep{2017:CohenFialkovBarkana}. This result inspired an array of attempted beyond-standard model explanations ranging from weakly-interacting dark matter models to enhanced radiation backgrounds \citep{2018:Barkana, 2018:MunozLoeb, 2018:FraserHektorHutsi,2018:PospelovPradlerRuderman,2019:BrandenbergerCyrShi,2019:MirochaFurlanetto}. However, while the size of the trough was successfully replicated in these models, its shape was more difficult to capture, with \cite{2022:MittalKulkarni} showing that this requires either highly unusual stellar populations or strong constraints on the cosmic star formation rate density, detectable by the James Webb Space Telescope. Additionally, the SARAS team published a non-detection of the EDGES signal in their data \citep{2022:SinghJishnuSubrahmanyan}. 

Typically, 21-cm foreground/signal separation relies on the spectral smoothness of the foregrounds relative to the 21-cm signal. The largest contribution to the 21-cm foregrounds comes from Galactic synchrotron emission, the temperature of which characteristically scales as a featureless power law in frequency \citep[$T_\mathrm{synch}\propto \nu^{-\gamma}$,][]{1999:ShaverWindhorstMadau}. The spectral index of this emission varies gradually throughout the sky, so any wide-beamed experiment averaging multiple sky regions into a single spectrum observes a power law with running terms. 

In order to leverage this smoothness, a single spectrum fit (SSF) can be performed on the measured sky signal with the sum of a smooth power law based function, and a physical or phenomenological 21-cm signal function. The trouble with this approach is that non-spectrally smooth systematic errors remaining in the data may be confused with the signal - an issue aggravated by the slight degeneracy between the foreground and signal functions typically used. Various re-analyses of the EDGES data that use this general methodology  \citep{2018:HillsKulkarniMeerburg,2019:SinghSubrahmanyan,2020:SimsPober} have suggested the possibility of a sinusoidal systematic in the data. Unfortunately, while these analyses indicate that the data favours the presence of a systematic, a wide variety of best-fit 21-cm signals, including the complete absence of a signal, are not significantly favoured over one another.

This presents a problem -- with the existence of unknown experimental systematics in the EDGES data, it may not be possible to be truly confident in a signal detection. It has been argued that the detection of the dipole of the cosmological 21-cm signal could help lift this ambiguity by acting as a verification of a true all-sky signal detection \citep{2017:Slosar,2018:Deshpande, 2024ApJ...964...21H,2024arXiv240913020M}, however this would require sensitivities well beyond the current generation of 21-cm experiments \citep{2024:IgnatovPritchardWu}. These issues demonstrate the importance of a clear understanding and control of any possible sources of systematic error in global 21-cm experiments, and motivates the requirement for multiple experiments with a wealth of alternative signal extraction pipelines to agree on a particular claimed detection. 

A major source of systematic error highlighted by \cite{2015:BernardiMcQuinnGreenhill} and \cite{2016:MozdzenBowmanMonsalve} is the interaction of the beam chromaticity with the spatial shape of the foregrounds, causing spectral oscillations larger than the expected amplitude of the 21-cm signal. This has led to the development of a number of foreground models that explicitly simulate both the chromaticity of the observing beam and the shape of the foregrounds. These include the Singular Value Decomposition (SVD)-based method of \cite{2018:TauscherRapettiBurns,2020:RapettiTauscherMirocha,2020b:TauscherRapettiBurns}; a framework that selects the optimal basis with which to fit the beam-weighted foregrounds through taking the SVD of many simulations of beam-weighted foreground observations, using beams of varying chromaticities and varying foreground instances. This method relies on the true foreground/beam chromaticity combination to exist within the simulated training set, but is shown to be fairly robust when applied to analysing mock data generated by the REACH experiment pipeline by \cite{2023:SaxenaMeerburgdeLeraAcedo}. 

Another such approach by \cite{2021:AnsteydeLeraAcedoHandley,2023:AnsteydeLeraAcedoHandley} (which we refer to as the N-regions model), takes a very different approach by modelling the foreground sky as a set of several regions, such that there is only a small variation in the power law index within each region. The foreground model scales the Haslam map \citep{2015:RemazeillesDickinsonBanday} at 408 MHz according to the power law index of each region, with the power law indices of the regions as the model parameters. However, any errors in the Haslam basemap bias the inference, as considered by \cite{2024:PaganoSimsLiu}, who show that introducing additional amplitude scaling parameters to the model makes it robust to basemap errors on the order of 10\%.

In this paper, we present a new signal extraction method based on methods used for CMB mapmaking. In the CMB literature, it is often necessary to turn large amounts of timeseries data taken by a CMB experiment into a \textit{map} of the sky, consisting of a temperature value for each pixel or a set of spherical harmonics of the sky \citep{2006:deOliveira-CostaTegmark}. This effectively compresses the data, allowing for easier manipulation and inference. Typically, a linear matrix observation equation is used to map from the compressed representation to the data taken by an experiment with a known observation strategy, beamfunction and noise characteristics. The inversion of this equation to obtain the compressed-space representation given the larger-space representation is carried out using either maximum-likelihood estimation \citep{1997:Tegmark} or MCMC methods such as Gibbs sampling \citep{2023:KeihanenSuur-UskiAndersen}. 

While the possibility of making maps of the sky spherical harmonics from interferometer visibilities has previously been discussed \citep{2014:ShawSigurdsonPen}, here we present an application of this framework to global 21-cm experiments. In this work, we demonstrate the ability for these experiments to estimate the low-order chromaticity-free spherical harmonic coefficients of the sky. This allows us to apply a standard single-spectrum log-polynomial and 21-cm signal model fit on the recovered sky monopole. In order to apply this mapmaking methodology to the global 21-cm field, we show that data taken by an array of many latitudinally-separated antennas must be combined with a correction to account for foreground modes on small angular scales. We show that the mapmaking method succeeds in inferring the 21-cm signal given error in the correction on the order of 10\%.

This paper is structured as follows; we present the simulations we will use for the 21-cm signal and foreground sky in \cref{sec:21-cm_theory}. We present the linear observation equation to represent the measurement of these skies by multiple chromatic 21-cm global experiments in \cref{sec:linear_formalism}. In \cref{sec:temp space fitting} we discuss beam chromatic foreground coupling, showing how the SSF fails to infer the 21-cm signal when the data is affected by beam chromaticity, even when a chromaticity correction is applied to the data. In \cref{sec:GLS}, we adapt the CMB mapmaking methodology to the 21-cm context in order to infer the chromaticity-free monopole of the data. We discuss potential improvements to the method and shortcomings in our analysis in \cref{sec:discussion}, and conclude in \cref{sec:conc}.

\section{Sky Modelling}
\label{sec:21-cm_theory}
\subsection{21-cm Signal}
The 21-cm signal is formed due to the change in temperature of 21-cm wavelength background radiation passing through neutral hydrogen (HI) clouds in the early universe. The background radiation temperature $T_R$ (normally taken to equal to the CMB temperature) couples to the hyperfine spin-flip transition of HI, which has an associated temperature of $T_S$, defined by the occupation density of the lower and upper relative spin states
\begin{equation}
    \frac{n_1}{n_0} = 3e^{-T_*/T_S} \approx 3\left(1-\frac{T_*}{ T_S}\right)\,.
    \label{eq:spintemp}
\end{equation}
Here, $T_*$ defines the temperature corresponding to the characteristic energy of the 21-cm transition, and the factor of 3 comes from the degeneracy of the states. The sky-averaged \textit{differential brightness temperature} $T_{21}$ (i.e. the 21-cm signal) measures the change in the temperature of the radiation background due to the 21-cm interaction. When the first stellar objects formed, they emitted Lyman-series photons into the IGM, coupling the spin temperature to the much cooler gas temperature via the Wouthuysen-Field effect \citep{1952:Wouthuysen,1958:Field}, and forming the largest-predicted absorption trough of the global 21-cm signal.

For simplicity, in this work we solely focus on the first-stellar formation absorption trough. We simulate this trough by fitting a Gaussian to the signal generated by the {\small ARES} package \citep{2014:Mirocha} running its default parameters. The fiducial global 21-cm temperature is therefore
\begin{equation}
    T_{21}(\nu;\boldsymbol{\theta}_{21}) = -A_{21}\exp\left[-\frac{(\nu-\nu_{21})^2}{2\Delta^2}\right] \,,
    \label{eq:monopole}
\end{equation}
where $\boldsymbol{\theta}_{21}=(A_{21},\,\nu_{21},\,\Delta)$, and the fit parameters are $A_\mathrm{21}=132.42$ mK, $\nu_\mathrm{21}=68.57$ MHz and $\Delta=9.399$ MHz.

\subsection{Foregrounds}
A variety of physically-motivated foreground models have been used for the purposes of demonstrating the effectiveness of signal extraction methods, investigating the sensitivity of experiments to the 21-cm signal, and as part of experimental calibration. A common strategy extrapolates a measured base map (commonly the Haslam map \citep{1982:HaslamSalterStoffel}) backwards in frequency using a frequency-independent spectral index. Both simulations where the index is angularly independent \citep{2020a:TauscherRapettiBurns,2018:BowmanRogersMonsalve,2017:MozdzenBowmanMonsalve,2019:MozdzenMaheshMonsalve} and angularly dependent \citep{2021:AnsteydeLeraAcedoHandley,2024:PaganoSimsLiu,2023:SaxenaMeerburgdeLeraAcedo,2023:SimsBowmanMahesh} have been used, which we respectively refer to as Basemap Extrapolation (BE) and 2-Basemap Extrapolation (2BE) models. These names are chosen to reflect the fact that angularly independent models spectrally scale a single base map with a fixed power law index, while angularly dependent models tend to fit a power law to each pixel of two base maps at different frequencies. Exceptions to the latter do exist however -- \cite{2024:PaganoSimsLiu}, for example, model small deviations in the power law index when extrapolating from a single basemap.

For the purposes of our work, it is not necessary for the foreground simulation we use to represent the true sky with perfect accuracy. Rather, we choose a simulation with sufficient complexity to demonstrate how the standard SSF method can fail. The angular variation of the power law index in the 2BE model greatly reduces the efficacy of a SSF to data affected by beam chromaticity, even when a simple Beam Factor Chromaticity Correction \citep[BFCC,][]{2023:SimsBowmanMahesh} is applied in order to remove the effects of chromatic beams. Therefore, we use the 2BE model for this work.

In order to implement this model, we fit power laws with no running to each pixel of two base maps, one at 230 MHz, and another at 408 MHz. We refer to these base maps as the pixel vectors $\mathbf{T}_{230}$ and $\mathbf{T}_{408}$ respectively, ordered in the {\small HEALPIX}\footnote{{\scriptsize HEALPIX} website -- \url{http://healpix.sourceforge.net}} RING ordering scheme. The model is given by
\begin{equation}
    T_{\mathrm{2BE},p}(\nu) = \left(T_{408,p} - T_\mathrm{cmb}\right)\left(\frac{\nu}{408\,\mathrm{MHz}}\right)^{-\gamma_p} + T_\mathrm{cmb} \,,
    \label{eq:GSMA}
\end{equation}
where $T_{408,p}$ is the temperature of the $p^\mathrm{th}$ pixel of the vector $\mathbf{T}_{408}$. The CMB temperature is subtracted from the base maps and then re-added as it is invariant with frequency. The set of power law indices are given by
\begin{equation}
    \gamma_p = 
    \frac{
        \log\left(
            \frac{T_{230,p} - T_\mathrm{cmb}}
            {T_{408,p} - T_\mathrm{cmb}}
        \right)}
    {\log(230/408)} \,.
    \label{eq:GSMA_idx}
\end{equation}
For convenience, throughout the rest of the paper we group the individual pixels of the above equations into vectors and drop the $p$ subscript notation, defining the pixel-space vectors $\T_\mathrm{2BE}$ and  $\boldsymbol\gamma$. The base maps are generated using the Global Sky Model \citep[GSM,][]{2008:deOliveira-CostaTegmarkGaensler}, a model that interpolates between many full and partial coverage radio sky surveys. The reason that we do not use the GSM itself to simulate the foreground sky is that, owing to the limited data available at low frequencies, the model displays a non-physical log-derivative of the foreground temperature with frequency in the 50-100 MHz band. 

In order to reduce the computational cost in this demonstrative paper, we opt to use lower resolution versions of the 2BE than the native GSM resolution of NSIDE=512 ($\sim 0.1\degr$), instead using NSIDE=32 ($\sim 1.8\degr$). We generate these by downgrading the resolution of the $\T_{230}$ and $\T_{408}$ basemaps used in \cref{eq:GSMA,eq:GSMA_idx} using the {\small UD\_GRADE} function of the {\small HEALPY} package. We discuss this further and justify our choice of resolution in \cref{sec:choice of Y}.

\section{Linear observation formalism}
\label{sec:linear_formalism}
In this section, we present the observation equation, which provides a matrix based mapping from a vector of spherical harmonic coefficients of the sky to a vector of observed, noisy timeseries data. We present the derivation of the general observation equation in \cref{sec:obs eq derivation}. In \cref{sec:obs eq application}, we tailor the observation equation to the 21-cm global experiment context, by presenting the structure and generation of the component matrices in the equation. The discussion of the conversion from the timeseries data vector back to the spherical harmonic coefficient vector using maximum-likelihood methods is left to \cref{sec:GLS}.

\subsection{Deriving the Observation Equation}
\label{sec:obs eq derivation}
A real function on a sphere, $f(\hat{\mathbf{n}})$, may be decomposed into a weighted sum of a set of real spherical harmonics\footnote{Since the functions we wish to express are real, we opt to use the real spherical harmonics.} $\{Y_{lm}(\hat{\mathbf{n}})\}$, and their corresponding real coefficients $\{a_{lm}\}$ as 
\begin{equation}
\begin{aligned}
    &f(\hat{\mathbf{n}}) = \sum_{l,m} a_{lm} Y_{lm}(\hat{\mathbf{n}}) \,,\\
    &Y_{lm}(\hat{\mathbf{n}}) = \begin{cases}
        \sqrt{2} P_{lm}(\theta) \sin m\phi  & m < 0  \\
         P_{lm}(\theta) & m=0 \\
         \sqrt{2} P_{lm}(\theta) \cos m\phi  & m > 0
    \end{cases}\;,
\end{aligned}
\label{eq:sh_decomposition}
\end{equation}
where $P_{lm}$ are the Legendre Polynomials. We may therefore express in this form the temperature of the sky $T_\mathrm{sky}(\hat{\mathbf{n}},\nu_i)$ in the direction $\hat{\n}$ and in a given frequency bin centred at $\nu_i$, the beamfunction $f_\mathrm{beam}(\hat{\mathbf{n}},\nu_i)$, and the beam-convolved temperature of the sky $T_\mathrm{conv}(\hat{\mathbf{n}},\nu_i)$. We take $a^{(i)}_{lm}$, $b^{(i)}_{lm}$ and $c^{(i)}_{lm}$ to be the respective spherical harmonic coefficients of each function, where $i$ is the frequency bin index. The beam-weighted sky temperature measured by a radiometer is given by the convolution of the beam function and the sky temperature, $T_\mathrm{conv}(\hat{\mathbf{n}},\nu_i)=(T_\mathrm{sky}*f_\mathrm{beam})(\hat{\mathbf{n}},\nu_i)$. It can be shown that, assuming an azimuthally-symmetric beam, this convolution on the sphere is equivalent to a multiplication of the spherical harmonic coefficients, with
\begin{equation}
\begin{aligned}
    &c^{(i)}_{lm} = \sqrt{\frac{4\pi}{2l+1}} a^{(i)}_{lm} b^{(i)}_{l0}\,, \\
    &T_\mathrm{conv}(\hat{\mathbf{n}},\nu_i) = \sum_{lm} \sqrt{\frac{4\pi}{2l+1}} a^{(i)}_{lm} b^{(i)}_{l0} Y_{lm}(\hat{\mathbf{n}}) \,, \label{eq:beam_convolved_alm}
\end{aligned}
\end{equation}
where we have simply substituted the first equation into \cref{eq:sh_decomposition} to obtain the convolved sky temperature as a function on a sphere in the second equation. We may write the expression for $T_\mathrm{conv}(\hat{\mathbf{n}},\nu_i)$ as the matrix equation
\begin{equation}
    \mathbf{T}^{(i)}_\mathrm{conv} = YB^{(i)}\a^{(i)}\,, \label{eq:beam conv temp}
\end{equation}
where each element of the vector $\mathbf{T}^{(i)}_\mathrm{conv}$ corresponds to a pixel direction $\hat{\mathbf{n}}_p$ for all pixels on the sky $p=\{1,\ldots,N_\mathrm{pix}\}$, $\a^{(i)}$ is a vector of length $N_{l\mathrm{max}}=(l_\mathrm{max}+1)^2$ containing the sky spherical harmonic coefficients in the order $ (a_{00}^{(i)},\,a_{1,-1}^{(i)},\,a_{1,0}^{(i)},\ldots,a_{l\mathrm{max},l\mathrm{max}}^{(i)})$, and $B^{(i)}$ is a diagonal matrix re-packaging of the non-zero spherical harmonic coefficients of the beam, featuring $(2l+1)$ copies of each $b^{(i)}_{l0}$ coefficient for each value of $l$, each weighted by the square root factor in \cref{eq:beam_convolved_alm}. The \textit{spherical harmonic matrix} elements are $Y_{pq}\equiv Y_{l_q m_q}(\hat{\mathbf{r}}_p)$, where the subscript $q$ denotes the same $(l,m)$ ordering as the spherical harmonic vector, whereas $\hat{\mathbf{r}}_p$ are the unit vectors corresponding to the pixel positions in the sky.

In order to make mock timeseries observations of the sky, one would choose a set of pointing vectors $\{\hat{\mathbf{n}}_t\}$ for $t=\{1,\ldots,N_t\}$, where $N_t$ is the number of timeseries data points, at which to evaluate the function $T_\mathrm{conv}(\hat{\mathbf{n}},\nu_i)$. In the matrix formalism, this is equivalent to the multiplication of $\mathbf{T}^{(i)}_\mathrm{conv}$ by a sparse \textit{pointing matrix} $P$ of shape $(N_{t},\, N_\mathrm{pix})$, which contains a row for each timeseries data point. Each row contains a single "1" at the index corresponding to the pixel at which $\{\hat{\mathbf{n}}_t\}$ points for that $t$ index. We write this in the matrix formalism as

\begin{equation}
    \mathbf{T}^{(i)}_\mathrm{obs} = P \mathbf{T}^{(i)}_\mathrm{conv} \,.
    \label{eq:Tobs Tconv}
\end{equation}
Note that a trivial (identity) pointing matrix would produce $\mathbf{T}^{(i)}_\mathrm{obs} =  \mathbf{T}^{(i)}_\mathrm{conv}$, corresponding to the case that all pixels on the sky are observed once.

Finally, timeseries data in the 21-cm field is commonly binned into a single, or a small number of time bins. Assuming that equal numbers of observed timeseries data points are averaged into each bin, we represent this through the multiplication of $\mathbf{T}^{(i)}_\mathrm{obs}$ by a $(N_{\tau},\, N_t)$ \textit{binning matrix} $G$, where $N_{\tau}$ is the number of bins. For our purposes, we always bin a fixed number of time points into a single bin and each time bin is the average of consecutive data only, therefore each row of the matrix contains a single string of $N_{\tau}/ N_t$, at the indexes corresponding to the times in that bin. A trivial (identity) binning matrix corresponds to the case that no binning is carried out. 

The observed data $\mathbf{d}^{(i)}$ is the raw data $\mathbf{T}^{(i)}_\mathrm{obs}$ with the addition of noise $\n^{(i)}$ in each time and frequency bin. Combining this with \cref{eq:beam conv temp,eq:Tobs Tconv} results in the observation equation
\begin{equation}
\begin{aligned}
    \mathbf{d}^{(i)} = GPYB^{(i)} \a^{(i)} + \n^{(i)} \, \equiv A^{(i)} \a^{(i)} + &\n^{(i)} \,, \,\\[5pt] &\mathrm{for} \, i\in \{1,\ldots,N_\mathrm{freq}\},\label{eq:cmb_mother}
    \end{aligned}
\end{equation}
where we have defined the $(N_\tau,N_{l\mathrm{max}})$ \textit{observation matrix} $A^{(i)}$ to represent the full mapping from sky spherical harmonic coefficients to noisy binned data. This set of $N_\mathrm{freq}$ matrix equations may conveniently be written as the single equation
\begin{equation}
    \mathbf{d}= \mathbf{GPYB} \a + \n \, \equiv \mathbf{A} \a + \n \,, \label{eq:cmb_mother block}
\end{equation}
by constructing block-diagonal matrices, denoted in bold font, out of the $N_\mathrm{freq}$-lots of individual matrices in \cref{eq:cmb_mother}. This multiplies the dimensionality of each component matrix by $N_\mathrm{freq}$, e.g. $\mathbf{A}$ is a $(N_\tau\, N_\mathrm{freq},N_{l\mathrm{max}} \,N_\mathrm{freq})$ matrix. Correspondingly, we have constructed the vectors $\d$, $\a$ and $\n$ by appending the $N_\mathrm{freq}$-lots of vectors $\d^{(i)}$, $\a^{(i)}$ and $\n^{(i)}$ together, resulting in e.g. the vector $\d$ of length $N_\tau \, N_\mathrm{freq}$. Note that this is for notational convenience alone - since all matrices used are block-diagonal, all matrix operations including inversion will result in the same block-diagonal structure. Therefore the portions of the vectors corresponding to different frequency components do not mix in with each other. All calculations may be carried out by splitting the block-diagonal matrices and vectors into their $N_\mathrm{freq}$ components, and recombining them after the calculation.

\subsection{Observation Matrices}
\label{sec:obs eq application}
The linear observation formalism in \cref{eq:cmb_mother block} is incredibly general, and in principle can represent the observation of the spherical harmonic coefficients of any sky by any azimuthally-symmetric beam with any observation strategy and any noise model, depending on the construction of the matrices and vectors used in \cref{eq:cmb_mother}. In this section, we present the choices we will use in this paper to simulate the observations of the 21-cm foregrounds and global signal by multiple latitudinally-separated drift-scan antennas with identical beam functions. The formalism and methodology used throughout the rest of the paper can readily be applied to multiple antennas with differing beams, as we note in \cref{sec:observation strategy}.

In \cref{sec:observation strategy} we will discuss the observation strategy encoded by the structure of the matrices $P$, $G$, and the noise profile $\mathbf{n}$. We discuss the antenna beam encoded by $\mathbf{B}$ in \cref{sec:choice of B}, and in \cref{sec:choice of Y} we discuss up to which spherical harmonic mode ($l_\mathrm{max}$) and up to which pixel resolution ($N_\mathrm{pix}$) to model in this work. 

\subsubsection{Observation strategy}
\label{sec:observation strategy}
In order to reconstruct spherical harmonic modes, a number of spatially-separated observations of the sky are required. We therefore simulate the observations of the 21-cm sky taken by a global antenna array composed of odd numbers of identical latitudinally-separated antennas. In all cases considered, a single antenna is placed at the equator and all other antennas are placed symmetrically North and South of it, with one every $26\degr$. For example, the antenna latitudes for the $N_\mathrm{ant}=5$ case are $52\degr \mathrm{N},26\degr \mathrm{N},0\degr,26\degr \mathrm{S}$ and $52\degr \mathrm{S}$. For simplicity, we place all antennas at longitude $0\degr$. 

Each antenna integrates continuously over the sidereal day, binning its observed temperature into timeseries data with $N_t=240$ time bins per day\footnote{In reality, the observed temperature in each bin is a time-average of the temperature observed as the beam drifts across the sky from the beginning to the end of the time bin. However, observations in time bins separated by $\lesssim 10$ min are well-approximated by instantaneous pointings \citep{2023:SimsBowmanMahesh}.}. We represent the pointings of a single antenna in the array with a pointing matrix of shape $(N_t,N_\mathrm{pix})$. Since the antennas simulated are identical, we construct a total pointing matrix by stacking the individual pointing matrices of each antenna vertically, forming a matrix $P$ of shape $(N_t\,N_\mathrm{ant},N_\mathrm{pix})$.

Since the mapmaking reconstruction requires spatially-resolved data, in \cref{sec:GLS} we do not further bin the timeseries data. This corresponds to the use of an identity binning matrix, with $N_\tau=N_t\,N_\mathrm{ant}$. In order to bin the timeseries data into the single spectrum required in \cref{sec:temp space results}, we use a binning matrix with $N_\tau=1$. This corresponds to a matrix of shape $(1,N_t)$, where all values are $1/N_t$.

We simulate an observation scenario where the entire antenna array is switched on at the same time, and is switched off after a period of time equal to the total integration time $t_\mathrm{tot}$. We take the total integration time to be 200 hours, corresponding roughly with the order of magnitude integration time of the EDGES experiment used for the claimed detection \citep{2018:BowmanRogersMonsalve}. The antennas observe in 1 MHz bins in the frequency range of 50 MHz to 100 MHz inclusive. 

In both binning cases, we compute the noise vector $\mathbf{n}^{(i)}$ for each frequency bin by sampling from a $N_\tau$-dimensional Normal distribution with mean zero and variance $\boldsymbol{\sigma}(\nu_i)^2$, where the variance is an $N_\tau$-dimensional vector calculated using the radiometer equation
\begin{equation}
    \boldsymbol{\sigma}(\nu_i) = \frac{\mathbf{d}^{(i)}}{\sqrt{ t_\mathrm{tot}\,\Delta \nu/N_\tau}} \,,
    \label{eq:radiometer_eq}
\end{equation}
where $\Delta \nu=$ 1 MHz is the bandwidth of a single frequency bin. Note that, since $N_\tau$ is different in \cref{sec:temp space fitting} to \cref{sec:GLS}, the noise in each frequency bin will differ across these sections.

\subsubsection{Antenna beam}
\label{sec:choice of B}
We assume a perfectly characterised, rotationally symmetric beam pointed at the zenith, given by
\begin{equation}
    f_\mathrm{beam}(\theta;\,w(\nu)) \propto 
    \begin{cases}
        \cos^2\left(\pi\,\theta/2 w\right) & \text{for } \theta < w(\nu) \, \\
        0 & \text{otherwise } 
    \end{cases}
\end{equation}
where $\theta$ is the angular deviation from the zenith and $w$ is the FWHM of the beam at a given frequency $\nu$. The conversion to spherical harmonic coefficients of the beam $b_{lm}^{(i)}$ is performed by evaluating the beam at each pixel of a {\small HEALPIX} map and multiplying the corresponding vector by $Y^{-1}$  (see \cref{sec:choice of Y}). Since the beam is rotationally symmetric, all $m\neq 0 $ modes are zero.

Throughout this work we use both achromatic and chromatic beams. Achromatic beams are modelled using $w=\mathrm{const}$. The spherical harmonic decomposition of a simple cosine-squared beam with $w=0.5\pi$ rad is zero for $l=\{4,6,8,\ldots\}$. The multiplication of the beam matrix with the sky spherical harmonics in \cref{eq:beam conv temp} would suppress these sky modes, leading to no information from these modes appearing in the data vector. Therefore, we choose to use the slightly lower width of $w=0.4\pi$ rad $=72\degr$.

To simulate beam chromaticity, we use a chromaticity profile following \cite{2020a:TauscherRapettiBurns}, which models the FWHM as a parabola in frequency, interpolating between two different end-point frequencies. Varyingly-chromatic beams are simulated by changing the curvature parameter of the parabola, $c$. While true beam chromaticity curves can be much more complex than this and feature higher frequency spectral oscillations,  \cite{2020a:TauscherRapettiBurns} show that this chromaticity profile already leads to false 21-cm signals being inferred when no 21-cm signal is present in the data.

\begin{figure}
    \centering
    \includegraphics[width=\columnwidth]{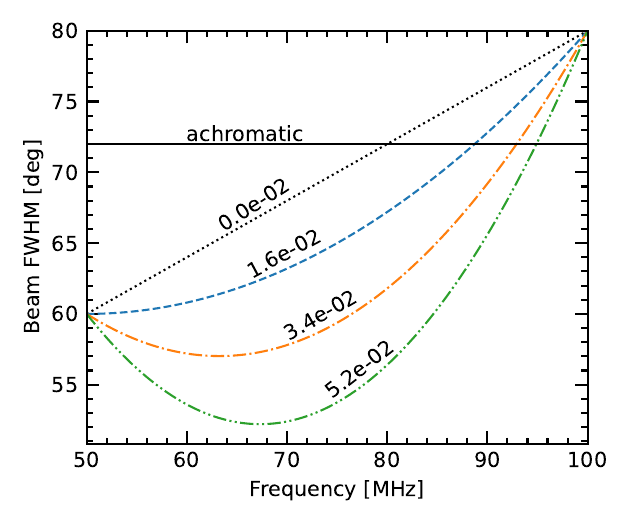}
    \caption{The FWHM of the cosine-squared beam as a function of frequency, for the achromatic case (bold line), as well as for three chromatic profiles we consider with varying values of the curvature parameter $c$, the value of which is labelled on each line. The FWHM corresponding to $c=0$ (dotted line) is also shown for reference.}
    \label{fig:fwhm}
\end{figure}

The spectrally-varying FWHM, $w(\nu)$, is given by
\begin{equation}
\begin{aligned}
    &w(\nu) = w_l(\nu) + c w_c(\nu) \\
    &w_l(\nu) =  (w_1-w_0)\nu/(\nu_1 - \nu_0) + 2 w_1 - w_0\\
    &w_c(\nu) = 0.5(\nu-\nu_1)(\nu-\nu_0)
\end{aligned}
\end{equation}
for the two end-point frequencies $\nu_0=50$ MHz and $\nu_1=100$ MHz with FWHM values $w_0$ and $w_1$. We use the values of $c$ used in \cite{2020a:TauscherRapettiBurns}, which produce the FWHM profiles shown in \cref{fig:fwhm}.

\subsubsection{Pixel resolution and $N_{l\mathrm{max}}$}
\label{sec:choice of Y}
In order to use \cref{eq:cmb_mother block} to generate mock observations of the 21-cm sky, the foreground and 21-cm skies must be represented as a vector of spherical harmonic coefficients $\a^{(i)}$ for each frequency, equal to the sum of the foreground spherical harmonics $\a^{(i)}_\mathrm{fg}$ and 21-cm monopole spherical harmonics $\a^{(i)}_{21}$. These spherical harmonic vectors are given by
\begin{equation}
    \begin{aligned}
        &\a_\mathrm{fg}^{(i)}=Y^{-1}\mathbf{T}^{(i)}_\mathrm{fg} \,,\\
        &\a^{(i)}_{21} = \sqrt{4\pi}\left(T_{21}(\nu_i;\boldsymbol{\theta}_{21}),0,\ldots,0\right) \,,
    \end{aligned}
\end{equation}
where we only include the monopole of the 21-cm signal, assuming that higher order multipoles of the signal are insignificant. $Y^{-1}$ is the \textit{pseudo-inverse} of the spherical harmonic matrix. The pseudo-inverse is defined as the matrix that solves the equation $Y\a_\mathrm{2BE}^{(i)}=\mathbf{T}^{(i)}_\mathrm{2BE}$. 

\smallskip

This definition is somewhat subtle. The spherical harmonic matrix $Y$ is a mapping from a truncated space of spherical harmonic components $\mathcal{H}$, representing modes up to and including $l_\mathrm{max}$, to a finite pixel space $\mathcal{P}$ of size $N_\mathrm{pix}$. Therefore, $Y^{-1}Y$ corresponds to the mapping $\mathcal{H} \rightarrow \mathcal{P} \rightarrow \mathcal{H}$, while $YY^{-1}$ is the mapping $\mathcal{P} \rightarrow \mathcal{H} \rightarrow \mathcal{P}$. If the intermediate space which is being transformed through is significantly compressed with respect to the original/target space, the overall transformation will be lossy. For instance, if $\mathcal{H}$ is significantly ``smaller'' than $\mathcal{P}$, then transforming a pixel map into a truncated spherical harmonic space with $Y^{-1}$ will lose information. Once the map is returned to pixel space with $Y$, it will be less detailed than the original pixel space map.

In this paper, we choose $\mathcal{H}$ and $\mathcal{P}$ such that $\mathcal{H}$ is compressed relative to $\mathcal{P}$. This choice is natural given the fiducial foregrounds we work with are defined in pixel space as a series of pixels with varying spectral index. The complexity of the foreground map lies in the variation of the spectral index for each pixel of the sky. This complexity is only well-captured by a large $\mathcal{P}$. Therefore, $YY^{-1} \neq \mathbb{I}_{N\mathrm{pix}\times N\mathrm{pix}}$, whereas $Y^{-1}Y=\mathbb{I}_{Nl\mathrm{max}\times Nl\mathrm{max}}$, as the latter represents a transformation from a compressed space of spherical harmonics, into a finite pixel space large enough to fully represent the information in the map, and back to the compressed spherical harmonic space. 

Given this choice, foreground map detail will be lost in the multiplication $\a_\mathrm{2BE}^{(i)}=Y^{-1}\mathbf{T}^{(i)}_\mathrm{2BE}$. However, since the antenna beam is very broad, the beam's spherical harmonic coefficients rapidly vanish for larger $l$ modes. The multiplication of $\a_\mathrm{2BE}^{(i)}$ by these modes results in the beam-convolved sky $\mathbf{T}^{(i)}_\mathrm{conv}$ being less sensitive to the higher-order spherical harmonics of the original sky.

In order to choose a large enough $l_\mathrm{max}$ to realistically represent the problem at hand while keeping computational cost low, we investigate at which $l$ mode the observed 2BE temperature contributions become smaller than the characteristic scale of the 21-cm signal. We perform beam-weighted observations of the 2BE sky for $l_\mathrm{max}=\{2, 4, \ldots, 32, 64\}$, and observe how much the resulting temperature changes for each doubling of $l_\mathrm{max}$. We seek to choose a value of $l_\mathrm{max}$ for which the effect of modelling to $2l_\mathrm{max}$ provides a smaller change than the magnitude of the 21-cm signal ($\sim 100$ mK). 

These mock observations are carried out with a 7-antenna array, binning the data into a single spectrum using $N_\tau=1$ as outlined in \cref{sec:observation strategy}, however we do not add noise to the data. For each value of $l_\mathrm{max}$, we compute the RMS difference between the observed data vector at 70 MHz and the data vector resulting from observing the sky truncated at $2l_\mathrm{max}$. This frequency is chosen to approximately coincide with the peak of our fiducial 21-cm signal. 

\begin{figure}
    \centering
    \includegraphics[width=\columnwidth]{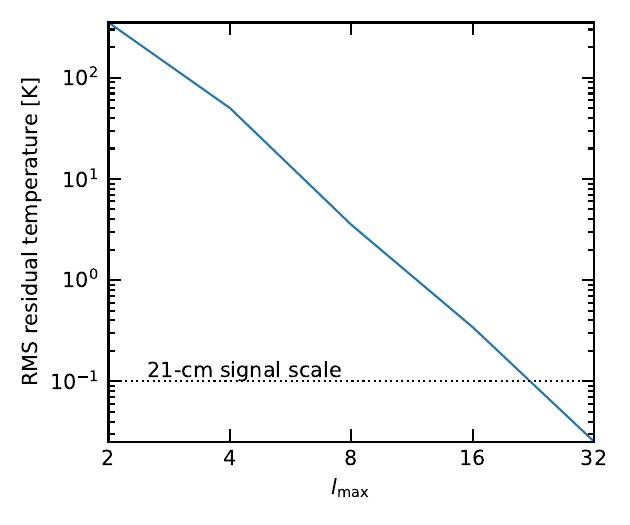}
    \caption{The RMS temperature difference between observations of the 2BE foreground sky truncated at  $l_\mathrm{max}$ and $2l_\mathrm{max}$. Noiseless mock observations at 70 MHz are used, generated with the standard observing strategy and binned into a single time bin. This RMS difference is compared to the approximate 21-cm signal scale (dotted line).}
    \label{fig:lmod_nside_investigation}
\end{figure}

The RMS differences are shown in \cref{fig:lmod_nside_investigation}, compared to the approximate scale of the 21-cm signal. The differences between $l_\mathrm{max}$ and $2l_\mathrm{max}$ approximately follow a power law, with roughly a power of 10 decrease for every doubling of $l_\mathrm{max}$. The residual RMS between $l_\mathrm{max}=32$ and $2l_\mathrm{max}=64$ is approximately an order of 5 smaller than the magnitude of the fiducial 21-cm monopole temperature. Following the trend of \cref{fig:lmod_nside_investigation}, the RMS difference between $l_\mathrm{max}=64$ and $2l_\mathrm{max}=128$ is expected to be an order of magnitude smaller than this. Therefore, the RMS error at 70 MHz resulting from working with a spherical harmonic space truncated at $l_\mathrm{max}=32$, as opposed to a non-truncated space, is also approximately an order of 5 smaller than the fiducial 21-cm monopole temperature. For the purposes of this work, we therefore use $l_\mathrm{max}=32$.

In order for the pixel space to adequately represent up to $l_\mathrm{max}$ modes, the length scale of each pixel must be greater than the size of the smallest angular scale represented by $l_\mathrm{max}$. This latter angular scale is commonly approximated as $\delta\theta_{SH} \sim 180\degr/l_\mathrm{max}$, as the $m= l$ and $m=-l$ modes divide the $2\pi$ rad diameter of the unit sphere into $2l$ sections. In the {\small HEALPIX} pixelisation scheme, {\small NSIDE} refers to the number of divisions along one side of the lowest-possible resolution {\small HEALPIX} pixel required to reach the desired resolution, with $N_\mathrm{pix} = 12\,\mathrm{NSIDE}^2$. The solid angle subtended by each equal-area pixel is $4\pi/N_\mathrm{pix}$. If the pixels are approximated as squares, the approximate length scale of a single pixel is therefore $\delta\theta_\mathrm{pix}=\sqrt(4\pi/12\,\mathrm{{\small NSIDE}^2})$ rad. The pixel width for {\small NSIDE}=32 corresponds to an angular scale approximately 3 times smaller than $l_\mathrm{max}=32$, so this choice is used for the rest of the paper.

\section{Beam Chromatic Foreground Coupling}
\label{sec:temp space fitting}
It is known that chromatic antenna beams couple angular structure in the sky to spectral structure in the measured data \citep{2015:BernardiMcQuinnGreenhill,2016:MozdzenBowmanMonsalve}. This has been previously addressed in the context of SSFs with a beam factor chromaticity correction (BFCC) based on the Haslam map by \cite{2017:MozdzenBowmanMonsalve,2019:MozdzenMaheshMonsalve}. This correction involves the calculation of a chromaticity-correction factor $C(\nu)$ for each pointing. The timeseries data for each pointing is then divided by the correction factor before being binned. The correction factor is given by
\begin{equation}
C(\nu) = \frac{\int  \mathrm{d}\mathbf{n}f_\mathrm{beam}(\mathbf{n};\nu) T_\mathrm{ref}(\mathbf{n};\nu) }{\int  \mathrm{d}\mathbf{n}f_\mathrm{beam}(\mathbf{n};\nu_\mathrm{ref}) T_\mathrm{ref}(\mathbf{n};\nu) }  \,,\label{eq:chrom corr}
\end{equation}
where $T_\mathrm{ref}$ is the reference foreground model, and $\nu_\mathrm{ref}$ is a reference frequency within the measured frequency range. If $T_\mathrm{ref}$ is equivalent to the true foregrounds, the BFCC results in corrected timeseries data at all frequencies which has effectively been measured by the beam shape at the reference frequency $f_\mathrm{beam}(\mathbf{n};\nu_\mathrm{ref})$, i.e. as if the experiment were achromatic. 

Since the true foreground sky is unknown, the reference foregrounds have been simulated as a BE model, using the Haslam map with a power law index of 2.5, has been used as the reference foregrounds in order to correct EDGES data, including the claimed EDGES detection \citep{2018:BowmanRogersMonsalve,2017:MozdzenBowmanMonsalve,2019:MozdzenMaheshMonsalve}. 

The BFCC has been shown to fail when the spatial \citep{2024:PaganoSimsLiu}  and spectral \citep{2023:SimsBowmanMahesh,2021:AnsteydeLeraAcedoHandley} structure of the reference foregrounds sufficiently differ from the true foregrounds, and when the beam shape is not perfectly characterised \citep{2022:SpinelliKyriakouBernardi}.  \cite{2023:SimsBowmanMahesh} have developed a \textit{BFCC data model} to fit BFCC corrected data without biasing the inferred 21-cm signal. This data model can account for errors in the spectral structure of the reference model, however it still requires perfect knowledge of the true foregrounds at the reference frequency, i.e. $T_\mathrm{ref}(\mathbf{n};\nu_\mathrm{ref})$.

In the form we will present in this paper, the mapmaking method is able to infer the 21-cm signal without bias in chromatic data. It is able to achieve this without requiring error-free reference foregrounds at any frequenc(y/ies). Rather, it requires an estimate of the mean and covariance of the foregrounds (see \cref{sec:missing modes corr,sec:alt foregrounds}). An extension to this method which may account for errors in the beam characterisation is left for further work.

\subsection{Beam Factor Chromaticity Correction}
\label{sec:temp space results}
In order to highlight the performance of the mapmaking method in the presence of chromatic beams, we benchmark it against the BFCC SSF. Casting \cref{eq:chrom corr} into a linear matrix formalism, the chromaticity correction vector is defined as
\begin{equation}
    \mathbf{C}^{(i)}= \frac{PYB^{(i)}\a^{(i)}_\mathrm{ref}}{PYB^{(i)}_\mathrm{ref}\a^{(i)}_\mathrm{ref}}  \,.\label{eq:chrom corr vec}
\end{equation}
Here, and throughout the rest of the paper, division of vectors is understood to be element-wise. $\a^{(i)}_\mathrm{ref}$ is the spherical harmonic vector of the scaled Haslam map and $B^{(i)}_\mathrm{ref}$ is the reference beam matrix. This is constructed such that all frequency blocks are a copy of the block $B^{(60\mathrm{MHz})}$, i.e. representing the beam at a reference frequency of 60 MHz. The chromaticity correction must be applied to the observed data before it is binned, giving the chromaticity-corrected data vector 
\begin{equation}
    \mathbf{d}_\mathrm{corr}^{(i)} = G \left[(PYB^{(i)} \a^{(i)} )/\mathbf{C}^{(i)}\right]+ \n^{(i)} \,.
\end{equation}
Note that here $G$ bins all data into a single spectrum, so $\d_\mathrm{corr}^{(i)}$ is a 1-element vector.

Data is generated according to the observation strategy outlined in \cref{sec:observation strategy} with a 7-antenna array. The foreground and 21-cm sky are observed in the case that all antenna beams are achromatic, and in two cases where all antenna beams are chromatic, with chromaticity curvatures $c=1.6\times10^{-2}$ and $c=3.4\times10^{-2}$. The BFCC-corrected single spectrum is fit with the function $\mathcal{M}(\nu;\boldsymbol\theta) = F(\nu;\boldsymbol\theta_\mathrm{fg}) + T_{21}(\nu;\boldsymbol\theta_{21}) + T_\mathrm{cmb}$, where $F$ is the log-polynomial function 
\begin{equation}
    F(\nu;\boldsymbol\theta_\mathrm{fg}) = \exp \left\{\sum_{\alpha=0}^{N_\mathrm{poly}-1} \theta_\mathrm{fg}^{(\alpha)} \left[\log \left(\nu/60\,\mathrm{MHz}\right)\right]^\alpha \right\} + T_\mathrm{cmb} \,.
    \label{eq:log_polynomial}
\end{equation}
Parameter inference is carried out using the MCMC package {\small EMCEE} \citep{2013:Foreman-MackeyHoggLang}.

In order to choose the optimal number of log-polynomial coefficients $N_\mathrm{poly}$ for each beam chromaticity value used, we compute the Bayesian Information Criterion (BIC) for each fit. The BIC is a measure used for model selection, featuring the two terms
\begin{equation}
    \mathrm{BIC} = N_\mathrm{dof} \log (N_\mathrm{data}) - 2 \log (\hat{L}) \,,
\end{equation}
where the first penalises the model for its complexity, and the second rewards it for describing the data well. $N_\mathrm{dof}=N_\mathrm{poly}+3$ is the number of model degrees of freedom and $N_\mathrm{data}$ is the number of temperature-space data points. The larger the number of polynomial coefficients, the larger the BIC. The second term is the negative log-likelihood evaluated at the maximum likelihood position. Models with a lower BIC are preferred, which is achieved with a higher maximum-likelihood value and a lower number of model parameters used. 

\begin{figure}
    \centering
    \includegraphics[width=\columnwidth]{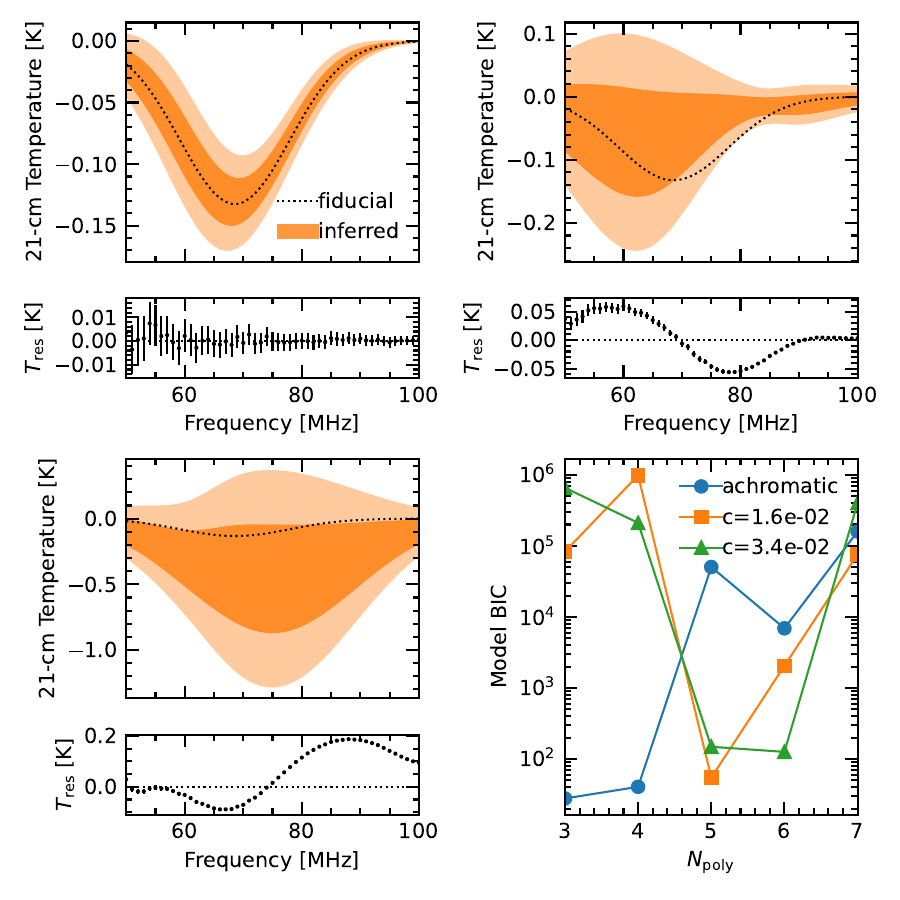}
    \caption{Results of the BFCC-corrected single-spectrum fit for data form a 7-antenna global array, collected with achromatic beams and fit with $N_\mathrm{poly}=3$ (top-left panel), a chromaticity curvature of $c=1.6\times10^{-2}$ and $N_\mathrm{poly}=5$ (top-right panel) and chromaticity curvature of $c=3.4\times10^{-2}$ and $N_\mathrm{poly}=6$ (bottom-left panel). For each panel, the upper sub-figure shows the coloured  $1\sigma$ and $2\sigma$ iso-probability contours of the functional posteriors of the inferred 21-cm signal, compared to the fiducial signal (dotted line). The lower sub-figure shows the full model residuals compared to zero (dotted line). The bottom-right panel shows the BIC for each chromaticity case as a function of $N_\mathrm{poly}$.}
    \label{fig:showcase_binwise}
\end{figure}

The results for all chromaticity cases are shown in \cref{fig:showcase_binwise}. The top-left, top-right and bottom-left panels correspond to the model inference for the achromatic case with $N_\mathrm{poly}=3$, the $c=1.6\times10^{-2}$  case with $N_\mathrm{poly}=5$ and the $c=3.4\times10^{-2}$ case with $N_\mathrm{poly}=6$ respectively. Each of these values of $N_\mathrm{poly}$ minimise the BIC for each chromaticity case. Each of these panels is made up of an upper and a lower sub-figure. The upper sub-figure shows the coloured  $1\sigma$ and $2\sigma$ iso-probability contours of the functional posteriors of the inferred 21-cm signal, compared to the fiducial signal. The iso-probability contours are computed by evaluating $T_{21}(\nu;\boldsymbol\theta_{21})$ at 10,000 points drawn from the marginalised 21-cm parameter posteriors. The lower sub-figure shows the full residuals to the single spectrum fit, i.e. the residuals after $\mathcal{M}(\nu;\boldsymbol\theta)$ evaluated at the mean posterior point is subtracted from the noisy data. 

The bottom-right panel shows the value of the BIC for each chromaticity case as a function of $N_\mathrm{poly}$. The ideal number of polynomial coefficients increases with increasing chromaticity, indicating that the residual spectral oscillations uncorrected for by the BFCC grow in complexity, requiring more polynomial terms to account for them. 

The iso-probability contours for the achromatic case clearly show that the BFCC SSF infers a 21-cm signal with the correct magnitude, width and position. The total model residuals are unpatterned and consistent with zero. Conversely, both chromatic cases feature ill-inferred 21-cm signals, with the $c=1.6\times10^{-2}$ case featuring a $1\sigma$ contour consistent with no signal, and the  $c=3.4\times10^{-2}$ case  featuring a very wide $1\sigma$ contour, encompassing 50 mK$\lesssim A_{21}\lesssim$ 850 mK. The residuals in both chromatic cases have spectral features on the order of magnitude of the 21-cm signal. 

It is likely that by further increasing the number of polynomial coefficients used for the chromatic fits, the foreground models would be able to fit the residual spectral features, producing total model residuals consistent with zero. This may even produce lower BIC values than those in \cref{fig:showcase_binwise}. However, increasing the number of coefficients used increases the degeneracy between the foreground model and the signal model, which would likely further broaden the inferred 21-cm iso-probability contours.

While our exact methodology differs, this result is qualitatively similar to \cite{2021:AnsteydeLeraAcedoHandley}, who find that single spectrum log-polynomial fits of chromatic observations of the 2BE feature spectral residuals larger than the scale of the 21-cm signal when the BFCC is applied. These results imply that when a 21-cm signal is injected into BFCC corrected data, and the data is fit with a log-polynomial and a 21-cm signal model, the inferred 21-cm signal will be biased. This agrees with our analysis in this section.

\section{The Mapmaking method}
\label{sec:GLS}
In \cref{sec:linear_formalism}, we presented a linear matrix formalism used to convert from spherical harmonic components of the sky, $\a$, to temperature spectra in multiple time bins observed by multiple chromatic drift-scan radiometers, $\d$. The inverse problem, that of calculating an estimate of the spherical harmonic coefficients of the sky $\hat{\a}$, given the data $\d$, is well-studied within the CMB field, using both maximum likelihood methods such as generalised least squares (GLS) \citep{1997:Tegmark} and Bayesian estimation methods \citep{2023:KeihanenSuur-UskiAndersen}. If it were possible to carry out this same estimation method within the 21-cm field using data from single-antenna radiometers, the estimated monopole temperature $\hat{T}_\mathrm{mon}(\nu_i)$ would contain the foreground monopole and the global 21-cm signal, uncontaminated by the interaction between beam chromaticity and foreground shape. The 21-cm signal could be inferred using the usual SSF of $\mathcal{M}$ to $\hat{T}_\mathrm{mon}(\nu_i)$. This signal extraction method would require no prior knowledge of the spatial or spectral structure of the foregrounds, while still being able to correct for beam-chromatic foreground coupling.

However, there are key differences between the nature of the data taken by CMB experiments and by 21-cm global experiments which prevent the direct application of the mapmaking method to global 21-cm inference. Within the CMB field, the inversion of \cref{eq:cmb_mother block} is typically an over-determined problem, with satellite-based CMB experiments having very narrow beams, capable of scanning every pixel of the sky multiple times. Within the 21-cm field, the same inversion is under-determined, as 21-cm global experiments consist of single zenith-pointing antenna instruments in fixed locations on the ground, with very broad beams. 

The lack of 21-cm global experiment sky coverage is a problem for the estimation of higher spherical harmonic modes. It has been shown previously that at least two latitudinally-separated radiometers must be used to constrain 21-cm dipole information \citep{2024:IgnatovPritchardWu}. As we show in \cref{sec:cmb gls}, the trend continues for higher order multipoles, with greater numbers of latitudinally-separated antennas required in order to probe a given spherical harmonic mode $l$. The inference of higher-order modes is additionally limited by the very broad beams typical of 21-cm experiments, which are insensitive to temperature variations on small angular scales.

The poor estimation of $l>0$ spherical harmonic modes can bias the accurate recovery of the $l=0$ mode, introducing errors far greater than the magnitude of the 21-cm signal (see \cref{sec:cmb gls}). Since the estimation of foreground modes above a certain spherical harmonic number is not possible using only the data from 21-cm global experiments, some prior information about the higher-order foreground modes must be assumed, in order to subtract their contribution from the measured data. The exact spatial and spectral structure of these modes is unknown, so we must rely on a correction based on the interpolation of existing low-frequency sky surveys. In order to use this correction while keeping the method robust in the case of errors in the surveys, we must use a foreground correction model which propagates the uncertainties of the low-frequency sky surveys on which it is based.

Standard sky models such as the GSM \citep{2008:deOliveira-CostaTegmarkGaensler} or the Low Frequency Sky Model \citep[LFSM,][]{2017:DowellTaylorSchinzel} use maximum-likelihood methods to interpolate in frequency between measured sky base maps, but they do not carry out error propagation. Additionally, these models suffer from large systematic errors arising from mean-zero calibration errors in the base maps used, which can be as high as 60\% \citep{2024:WilenskyIrfanBull}. 

The Bayesian Global Sky Model \citep[B-GSM,][]{2025:CarterHandleyAshdown} has been developed to address these issues by using global experiment data to calibrate the temperature scales of the base maps used in the model. The parameter inference is carried out using nested sampling \citep{2004:Skilling}, allowing for error propagation. The future application of the B-GSM to real data will provide a well-calibrated foreground model which propagates uncertainties. In anticipation of this, for the purposes of this paper we will construct a simple yet instructive stochastic foreground model which simulates noise in each pixel without calibration errors.

We present a the Generalised Least Squares (GLS) method used in the CMB field in \cref{sec:cmb gls} in order to investigate the number of modes which may be inferred with data taken by a global 21-cm antenna array without biasing the inferred monopole mode. \cref{sec:missing modes corr} extends the GLS method to include a correction for higher order foreground modes. In \cref{sec:alt foregrounds} we present the stochastic foreground model in order to generate this foreground correction. Finally, in \cref{sec:ML results} we apply the resulting GLS-based 21-cm mapmaking method to data taken with chromatic instruments.

\subsection{Reconstructing Sky Multipoles}
\label{sec:cmb gls}
We introduce apostrophe notation to denote vectors and matrices featuring a spherical harmonic vector index which is truncated at $l_\mathrm{mod}$. E.g. the full vector across all frequency bins $\a$ is composed of the appending of the set of vectors $\{\a^{(i)}\}$, each of which is $N_{l\mathrm{max}}$ long. The corresponding truncated vector $\a'$ is composed of the set of vectors $\{\a'^{(i)}\}$, each of which is $N_{l\mathrm{mod}}$ long, featuring only spherical harmonic coefficients up to $l_\mathrm{mod}$. 

For matrices, we similarly define the apostrophe notation as being a truncation imposed on all spherical harmonic indices of the matrix. E.g. $\A$ is truncated to $\A'$, a block-diagonal matrix composed of the blocks $A'^{(i)}$. Each of these blocks is made up of the components of the full matrix $A^{(i)}_{\alpha\beta}$, but the spherical harmonic index $\beta$ is constrained to the values $\beta \in  \{1,\ldots,N_{l\mathrm{mod}}\}$. For matrices with two spherical harmonic indices, such as $\mathbf{B}$, the frequency blocks will be made up of the components $B^{(i)}_{\alpha\beta}$ where both $\alpha$ and $\beta$ are constrained up to $N_{l\mathrm{mod}}$.

\begin{figure}
    \hspace{0.5cm}
    \includegraphics[width=\columnwidth]{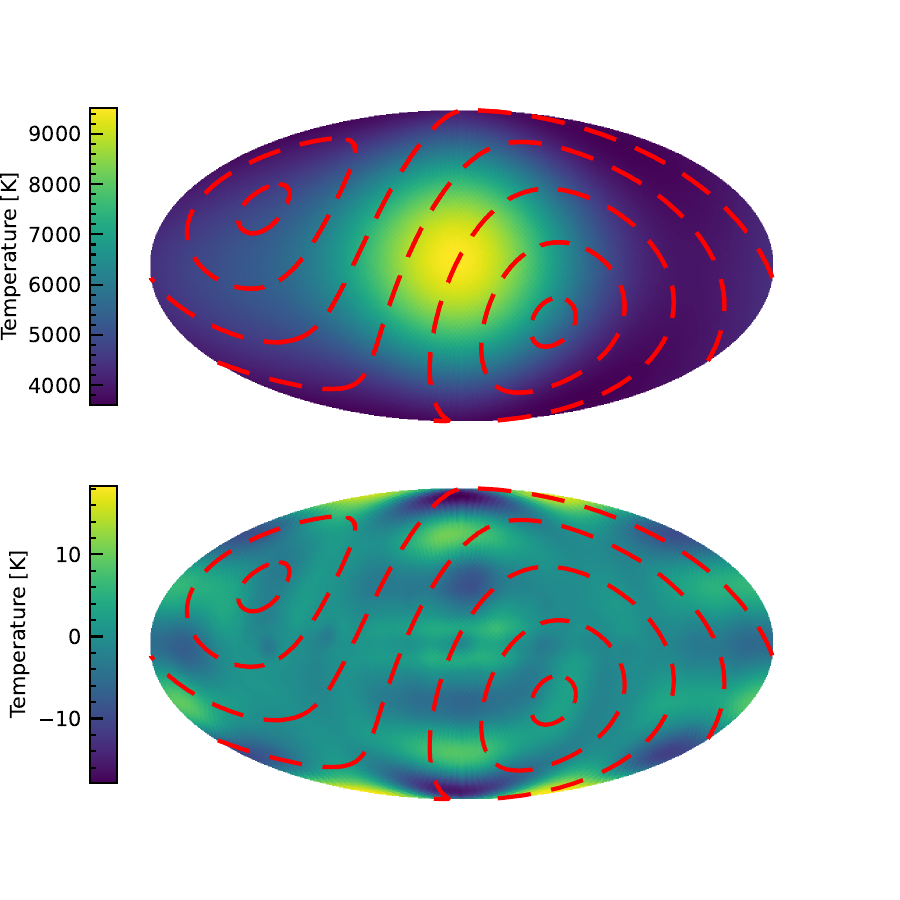}
\caption{Mollweide projection of the beam-convolved foreground sky at 60 MHz in Galactic coordinates for $l\leq 5$ (top) and $5 < l \leq 32$ (bottom). Overlaid are the drift-scan sky tracks of 7 antennas, equally spaced in latitude every $26\degr$ (red dashed lines).}
    \label{fig:skytrack_maps}
\end{figure}

We observe foregrounds without a 21-cm signal truncated at the orders $l_\mathrm{mod}=\{0,1,\ldots,12\}$ using the observation equation
\begin{equation}
    \d = \A' \a' + \n
    \label{eq:truncated obs eqn}
\end{equation}
where $\a'=\a'_\mathrm{fg}$. \cref{fig:skytrack_maps} (top) shows an example of the truncated beam-convolved foregrounds observed at 70 MHz with $l_\mathrm{mod}=5$. For each truncation order, three observation strategies are employed, featuring 3, 5 and 7-antenna arrays placed symmetrically about the equator (see \cref{sec:observation strategy}). The estimate of the sky spherical harmonic vector $\hat{\a}'$ is obtained using the Generalised Least-Squares (GLS) method. This is the optimal maximum-likelihood solution to the inversion of \cref{eq:truncated obs eqn}, derived from the minimization of the $\chi^2$ function \citep{1997:Tegmark}
\begin{equation}
    \chi^2 = (\d - \mathbf{A}'\a')^T\mathbf{C}^{-1}(\d - \mathbf{A}'\a') \,,
\end{equation}
where $\mathbf{C}=\mathbf{N}$ is the noise covariance matrix. Minimizing the $\chi^2$ function results in the maximum-likelihood estimate
\begin{equation}
\begin{aligned}
    &\hat{\a}'=\mathbf{W}'\d \,, \\
	&\mathbf{W}' \equiv [\mathbf{A}'^T\mathbf{C}^{-1}\mathbf{A}']^{-1}\mathbf{A}'^T\mathbf{C}^{-1} \equiv \mathbf{C}'_a \mathbf{A}'^T\mathbf{C}^{-1} \,,
    \label{eq:gen_soln}
   \end{aligned}
\end{equation}
where $\mathbf{C}'_a$ is the covariance matrix of the estimated spherical harmonic coefficients.

\begin{figure}
    \centering
    \includegraphics[width=\columnwidth]{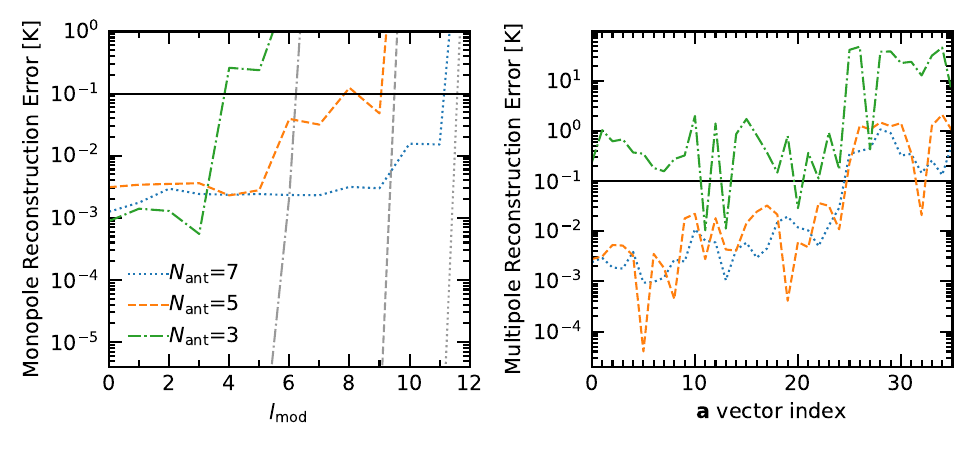}
    \caption{\textit{Left:} the absolute reconstruction error of the foreground monopole at 70 MHz, when the observed foregrounds are truncated at $l_\mathrm{mod}$ and a maximum likelihood reconstruction of order $l_\mathrm{mod}$ is used. Shown are the results for estimation based on data taken from 3 (dash-dot), 5 (dashed) and 7 (dotted) latitudinally-separated antennas. In grey are the corresponding monopole reconstruction errors for noise-free data. \textit{Right:} the absolute reconstruction error of the foreground spherical harmonic vector at 70 MHz, when the observed foregrounds are truncated at $l_\mathrm{mod}=5$ and a maximum likelihood reconstruction of order $l_\mathrm{mod}=5$ is used. In both figures, the approximate magnitude of the fiducial 21-cm signal at 70 MHz is shown (black line).}
    \label{fig:monopole_reconstruction_err}
\end{figure}

For each truncation order, the absolute error of the estimated monopole temperatures at 70 MHz are shown in \cref{fig:monopole_reconstruction_err} (left), compared to the approximate magnitude of the 21-cm signal. The absolute error is given by $|\hat{a}^{(70\mathrm{MHz})}_{0} - a^{(70\mathrm{MHz})}_{0}|/\sqrt{4\pi}$, where the division by $\sqrt{4\pi}$ converts the spherical harmonic coefficients to temperatures. The error in the estimated monopole temperature for low truncation orders approximately plateaus at the noise temperature at 70 MHz of $\sim 2$ mK for all observation strategies. For each number of antennas, the reconstruction error quickly grows in magnitude past a certain truncation order, with the reconstruction error from data taken with fewer antennas diverging for lower orders of $l_\mathrm{mod}$ than data from more antennas. 

For each number of antennas, \cref{fig:monopole_reconstruction_err} (left) also shows the monopole estimate error for noise-free data. Each noisy case features a reconstruction error which begins to increase in magnitude before the corresponding noise-free case does. This indicates that noise has a significant impact on the reconstruction accuracy, but the large divergences (above $\sim 10^3$ mK) coincide across the noise-free and noisy cases. This reveals the limit at which an array of 21-cm antennas with broad beams can no longer constrain the monopole mode.

While the monopole mode is well-constrained, higher order modes may not be. \cref{fig:monopole_reconstruction_err} (right) shows the absolute error of the estimated vector $\hat{\a}'^{(70\mathrm{MHz})}$ for the truncation order $l_\mathrm{mod}=5$. This error is given by $|\hat{\a}'^{(70\mathrm{MHz})} - \a'^{(70\mathrm{MHz})}|/\sqrt{4\pi}$. At this truncation order, \cref{fig:monopole_reconstruction_err} (left) shows that the error on the absolute monopole reconstruction error for the $N_\mathrm{ant}=3$ case is greater than the magnitude of the 21-cm signal, while the $N_\mathrm{ant}=5$ and $N_\mathrm{ant}=7$ cases feature monopole reconstruction errors on the order of $2$ mK.  \cref{fig:monopole_reconstruction_err} (right) shows that the reconstruction error of $l>0$ modes can be much greater than the reconstruction error of the $l=0$ mode, with both the $N_\mathrm{ant}=5$ and $N_\mathrm{ant}=7$ cases featuring high-order modes with reconstruction errors much greater than the 21-cm signal. Since the monopole mode itself is well constrained, these ill-constrained higher order modes do not affect the analysis.

Given the insensitivity of 21-cm global antenna arrays to higher order modes, an obvious question is whether a truncated maximum likelihood estimation of the spherical harmonics, applied to mock data of non-truncated foregrounds, would be able to recover the monopole temperature with sufficient accuracy. Testing this for the truncation orders $0\leq l_\mathrm{mod}\leq 12$, we find that the presence of unmodelled modes significantly biases the monopole reconstruction, resulting in $\gtrsim 10^3$ mK errors. 

These results roughly indicate that a 7-antenna array would require a foreground correction for modes above $l_\mathrm{mod}\approx10$. However, given an imperfect foreground correction, residual contributions from $l>l_\mathrm{mod}$ modes will be present in the data. This makes the sensitivity estimates shown in \cref{fig:monopole_reconstruction_err} (left) a best-case scenario, as the reconstruction of the monopole may be biased by the presence of these foreground residuals. For the rest of the paper, we therefore use a maximum likelihood reconstruction truncated conservatively at $l_\mathrm{mod}=5$ for data taken by 7 antennas, in order to ensure that the monopole reconstruction will be as unaffected as possible by poorly corrected foreground modes.

\subsection{Missing modes correction}
\label{sec:missing modes corr}
We include the contribution of the modes not constrainable by 21-cm global experiments in the model by splitting \cref{eq:cmb_mother block} into low-order modes truncated at $N_{l\mathrm{mod}}$ and higher-order modes truncated between $N_{l\mathrm{mod}}$ and  $N_{l\mathrm{max}}$. The observation equation becomes
\begin{equation}
    \d = \A'\a' + \A'' \a'' + \n \,, \label{eq:split obs eqn I}
\end{equation}
where the double-apostrophe notation denotes matrices and vectors that have had their spherical harmonic index truncated to values $\{N_{l\mathrm{mod}},\ldots,N_{l\mathrm{max}}\}$, analogously to the single-apostrophe notation used in \cref{sec:cmb gls}. \cref{fig:skytrack_maps} (bottom) shows the beam-weighted higher-order modes of the 2BE model for $l_\mathrm{mod}=5$. These higher modes are further divided into the mean of the foreground correction $\boldsymbol{\mu}''\equiv \langle \a''  \rangle$, and the variation of a particular realisation of the foreground correction from the mean $\boldsymbol\Delta''$, such that $\langle\boldsymbol\Delta''\rangle=0$. This is not to be confused with $\Delta$, the 21-cm signal width parameter. Substituting this into \cref{eq:split obs eqn I},
\begin{equation}
    \mathbf{d} - \A'' \boldsymbol\mu'' = \A'\a' + \A''\boldsymbol\Delta''+\n\,. \label{eq:split obs eqn II}
\end{equation}
The left-hand side of \cref{eq:split obs eqn II} corresponds to the foreground correction; the subtraction of the best estimate of higher mode foreground contributions to the data. The term $\A''\boldsymbol\Delta''$ represents the contribution to the data from imperfectly subtracted higher-order modes. $\A'\a'$ are the estimated modes, to be inferred using the GLS method, applied by minimising the chi-squared function 
\begin{equation}
    \chi^2 = \left[\left(\mathbf{d} - \A''\boldsymbol\mu''\right)-\A'\a' \right]^T \mathbf{C}^{-1} \left[\left(\mathbf{d} - \A''\boldsymbol\mu''\right)-\A'\a'\right] \,, \label{eq:split obs chi-sq}
\end{equation}
where now the covariance matrix is $\mathbf{C} = \mathbf{S} + \mathbf{N}$, where $\mathbf{S} \equiv \A'' \mathbf{C}''_a\A''^T \equiv \A''\langle \boldsymbol\Delta''\boldsymbol\Delta''^T\rangle\A''^T $ is the covariance in the data due to the imperfect higher-order mode foreground correction. The estimate of $\a'$ resulting from the minimisation of \cref{eq:split obs chi-sq} is
\begin{equation}
    \hat{\a}' = \mathbf{W}'(\d-\A''\boldsymbol\mu'') \,, \label{eq:specific soln}
\end{equation}
where $\mathbf{W}'$ is defined identically to \cref{eq:gen_soln}, where now the total covariance matrix $\mathbf{C}$ includes the corrected mode covariance term.

Minimising the chi-squared is the optimum solution to the estimation of $\a'$ if both $\n$ and $\boldsymbol\Delta''$ are Gaussian. This is true for the noise, both by construction and in reality. Non-Gaussianity may be expected in a realistic model of the foregrounds, which we discuss further in \cref{sec:discussion_gauss}.

\subsection{Stochastic Foreground Model}
\label{sec:alt foregrounds}
Thus far the 2BE has been used as the fiducial sky model. We extend this model by simulating uncertainty in the power law index of the 2BE model. This model will be referred to as the Stochastic 2-Basemap Extrapolation (S2BE) model. From now on, the fiducial sky will be taken to be a realisation of the S2BE, while the mean of the S2BE model will be used as the imperfect foreground correction.

The S2BE model adds independent Gaussian variation to the power law index of all pixels in \cref{eq:GSMA}, with $\boldsymbol{\gamma} \rightarrow \boldsymbol{\gamma} + \boldsymbol{\delta}$, where $\delta_p \sim G(\mu=0, \sigma=\sigma_\delta)$ for all $p\in \{1,\ldots,N_\mathrm{pix}\}$. Each time the model is run, it generates a new set of power law indices corresponding to a new realisation of the model, $\mathbf{T}_\mathrm{S2BE}(\boldsymbol\delta)$, given by the $N_\mathrm{freq}$-lots of $N_\mathrm{pix}$ vectors
\begin{equation}
    \mathbf{T}_\mathrm{S2BE}^{(i)}(\boldsymbol\delta) = \left(\mathbf{T}_{408} - T_\mathrm{cmb}\right)\left(\frac{\nu_i}{408\,\mathrm{MHz}}\right)^{-(\boldsymbol\gamma + \boldsymbol{\delta})} + T_\mathrm{cmb} \,.
    \label{eq:GSMA stoch}
\end{equation}
The spherical harmonic transform of this is denoted as $\a_\mathrm{S2BE}\equiv\mathbf{Y}^{-1}\T_\mathrm{S2BE}$. The mean of the model $\T_\mu\equiv\langle \T_\mathrm{S2BE} \rangle$, where the angle brackets denote the mean taken over multiple realisations of $\boldsymbol\delta$, is derived in \cref{apdx:mean_corr} as
\begin{equation}
   \T^{(i)}_\mu = \left( \mathbf{T}_{\mathrm{2BE}}^{(i)} - T_\mathrm{cmb}  \right) \exp\left[\sigma_T^2(\nu_i)/2\right]+ T_\mathrm{cmb} \,,
    \label{eq:GSMA stoch mean}
\end{equation}
where $\sigma_T(\nu)\equiv \sigma_\delta \log (408/\nu)$. The corresponding mean in spherical harmonic space is $\langle \a_\mathrm{S2BE}\rangle\equiv\boldsymbol{\mu}=\mathbf{Y}^{-1}\T_\mu$, while the mean truncated to modes higher than $l_\mathrm{mod}$ is $\boldsymbol{\mu}''=\mathbf{Y}''^{-1}\T_\mu$. Also derived in \cref{apdx:mean_corr} is the covariance matrix $\mathbf{C}_T$
\begin{equation}
\begin{aligned}
    &\mathbf{C}_{T} \equiv \langle \T_\mathrm{S2BE}\T_\mathrm{S2BE}^T\rangle -  \T_\mu\T_\mu^T \,,\\
    &C^{(i)}_{T,pp}=\left( T_{\mathrm{2BE},p}^{(i)} - T_\mathrm{cmb}  \right)^2 \left\{\exp\left[2\sigma_T^2(\nu_i)\right] - \exp\left[\sigma_T^2(\nu_i)\right]\right\}\,, 
\end{aligned}\label{eq:temp covar}
\end{equation}
the non-zero components of which are given by the second equality of \cref{eq:temp covar}. $\mathbf{C}_T$  is diagonal as the error in each pixel is independent. The covariance of the corresponding spherical harmonic coefficients $\a_\mathrm{S2BE}$ is given by $\mathbf{C}_a = \mathbf{Y}^{-1}\mathbf{C}_T (\mathbf{Y}^{-1})^T$, while the covariance of the truncated modes $\a''_\mathrm{S2BE}$ is $\mathbf{C}_a''=\mathbf{Y}''^{-1}\mathbf{C}_T(\mathbf{Y}''^{-1})^T$, also shown in \cref{apdx:mean_corr}.

Combining \cref{eq:GSMA,eq:GSMA stoch,eq:GSMA stoch mean}, it can be shown that the fractional deviation of each pixel of an instance of the S2BE model from the mean is
\begin{equation}
\begin{aligned}
    \frac{T^{(i)}_{\mathrm{S2BE,}p}(\delta_p) -  T_{\mu,p}^{(i)}}{ T_{\mu,p}^{(i)}} &= \frac{\left(\nu_i/408\right)^{-\delta_p} - \exp\left[\sigma_T^2(\nu_i)/2\right]}{ 
\exp\left[\sigma_T^2(\nu_i)/2\right] + T_\mathrm{cmb}/(T_{\mathrm{2BE},p}^{(i)}-T_\mathrm{cmb})}\\
    &\approx \delta_p \log \left(408/\nu_i\right) \,,
    \label{eq:approx_frac_err}
\end{aligned}
\end{equation}
where the approximate equality is computed in the limit of small $\sigma_\delta$ (and therefore small $\sigma_T$ and small $\boldsymbol{\delta}$), and by assuming that $\T_\mathrm{2BE} \gg T_\mathrm{cmb}$ for all pixels at low frequencies. Since $\delta_p$ is Gaussian-distributed, this result shows that each pixel of $\T^{(i)}_\mathrm{S2BE}(\boldsymbol\delta)$ is drawn from a roughly Gaussian distribution centred at $\T^{(i)}_{\mu}$, with an approximate width of $\sigma_T(\nu_i)\T^{(i)}_{\mu}$. In this Gaussian approximation, the S2BE model features independent fractional pixel errors at a percentage level specified by $\sigma_T$. Therefore, $\sigma_T$ is the fractional standard deviation of the S2BE model.

By setting $\sigma_\delta$, one chooses the value of $\sigma_T(\nu)$. As we will use the S2BE model's mean as an imperfect correction to $l>l_\mathrm{mod}$ foreground modes, $\sigma_T$ is referred to as the percentage error of the foreground correction. Since this percentage error varies with frequency as $\sigma_T$ is a function of frequency, percentage errors of the foreground correction are quoted at the reference frequency of $70$ MHz. For example, $\sigma_\delta = 0.057$ and 0.11 correspond roughly to $\sigma_T(70\,\mathrm{MHz})=0.1$ and 0.2, i.e. 10\% and 20\% errors respectively.

\begin{figure}
    \centering
    \includegraphics[width=\columnwidth]{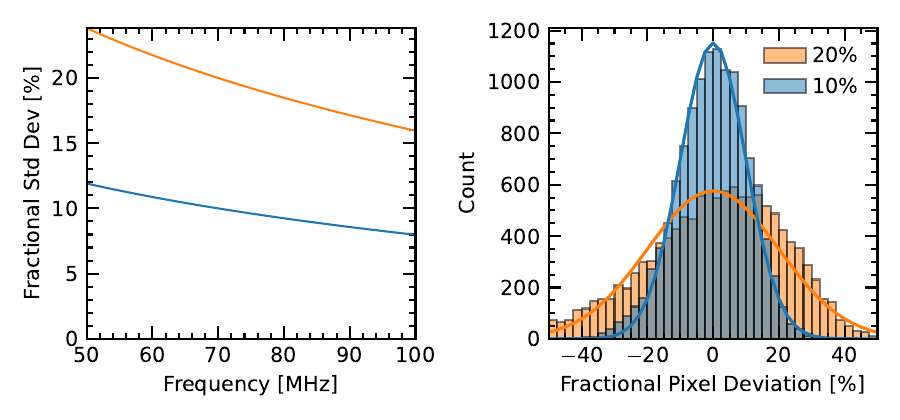}
    \caption{{Left:} the fractional standard deviation of the S2BE model in the Gaussian approximation. The lines correspond to a percentage error of 10\% and 20\% (bottom to top) at 70 MHz. \textit{Right:} percentage deviation from the mean of each pixel of a realisation of the S2BE model at 70 MHz, given the same reference distribution widths.}
    \label{fig:basemap_errs}
\end{figure}

\cref{fig:basemap_errs} (left) shows the approximate fractional pixel standard deviation as a function of frequency (i.e. $\sigma_T(\nu)\times 100\%$) for 10\% and 20\% percentage errors. \cref{fig:basemap_errs} (right) shows the true percentage deviation at 70 MHz of each pixel of a S2BE instance from the mean $\T_{\mu}(70\,\mathrm{MHz})$ for these two cases, i.e. the LHS of \cref{eq:approx_frac_err}. These are compared to mean-zero Gaussians with width 10\% and 20\% respectively. The distributions are well-approximated by the reference Gaussians, especially in the smaller $\sigma_\delta$ (10\% error) case.

\subsection{Results}
\label{sec:ML results}
The mapmaking method is applied to noisy mock data  generated using \cref{eq:cmb_mother block} with $\a = \a_\mathrm{S2BE} + \a_{21}$ in the case of achromatic beams in \cref{sec:non chrom ML results}, and chromatic beams with varying chromaticities in \cref{sec:chrom ML results}. The low-order modes are estimated using \cref{eq:specific soln} with $\boldsymbol\mu''=\mathbf{Y}''^{-1}\T_\mu$, and the covariance correction $\mathbf{S} = \A''\mathbf{Y}''^{-1}\mathbf{C}_T(\mathbf{Y}''^{-1})^T\A''^T$, with $\T_\mu$ and $\mathbf{C}_T$ corresponding to foreground corrections with varying percentage errors. For all cases presented, the recovered monopole was fit with the function $\mathcal{M} = F + T_{21}$ for the range of polynomial orders $N_\mathrm{poly}\in\{3,\ldots,7\}$. Only the inference results with polynomial orders which minimised the BIC are shown, which were found to be $N_\mathrm{poly}=3$ for all cases except the final one in \cref{sec:chrom ML results}, where the BIC preferred $N_\mathrm{poly}=4$.

\subsubsection{Achromatic case}
\label{sec:non chrom ML results}
\begin{figure}
    \centering
    \includegraphics[width=\columnwidth]{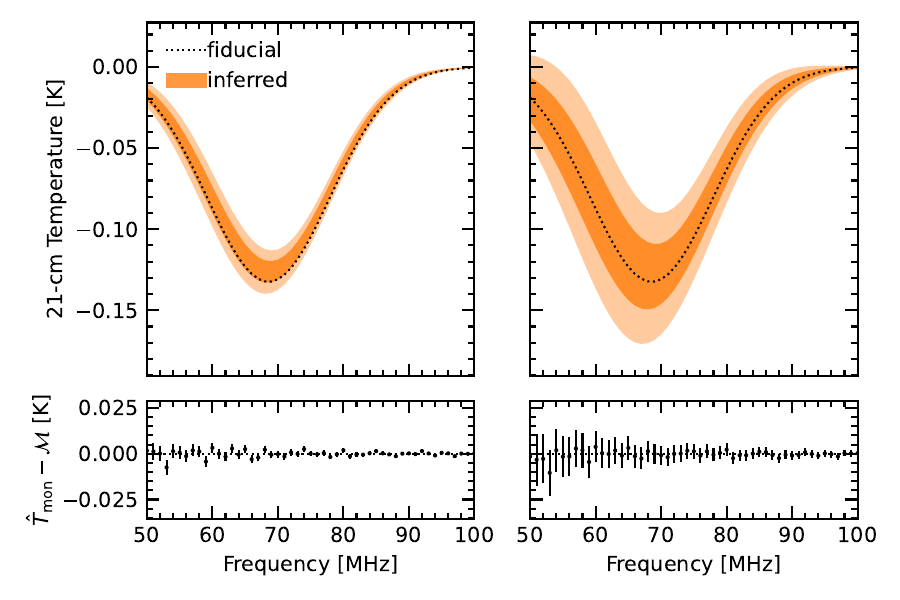}
    \caption{Results from the mapmaking method for data collected with an achromatic beam, where the contribution from modes above $l_\mathrm{mod}=5$ have been corrected for perfectly with $\sigma_T(70\,\mathrm{MHz})=0\%$ (left) and imperfectly with $\sigma_T(70\,\mathrm{MHz})=10\%$ (right). For both scenarios, an $N_\mathrm{poly}=3$ log-polynomial has been used to fit the foregrounds in the recovered monopole. The top subplots show the coloured $1\sigma$ and $2\sigma$ iso-probability contours of the functional posteriors of the 21-cm all-sky signal, compared to the fiducial signal (dotted line). The bottom subplots show the residuals to the MCMC fit of the model $\mathcal{M}$ to the maximum-likelihood recovered monopole temperature. The error bars indicate the uncertainty of the maximum-likelihood reconstruction of the monopole temperature.}
    \label{fig:ml_Nant<7>_Npoly<3>_achrom_idx<0>andNant<7>_Npoly<3>_achrom_idx<10>}
\end{figure}

\cref{fig:ml_Nant<7>_Npoly<3>_achrom_idx<0>andNant<7>_Npoly<3>_achrom_idx<10>} (left) shows the inferred 21-cm signal iso-probability contours for data taken with achromatic beams and a perfect foreground correction with $0\%$ error. The signal is very well constrained, with a $1\sigma$ width of $\sim 5$ mK at 70 MHz, and infers the fiducial signal to $1\sigma$. The estimated monopole temperature minus the fitted model $\mathcal{M}$ is shown in the lower panel, with error bars corresponding to the error on the estimate, computed as the square root of the monopole entries of the matrix $\mathbf{C}_a$, i.e. $\sqrt{C_{a,00}^{(i)}}$ for $i\in\{1,\ldots,N_\mathrm{freq}\}$. $\mathcal{M}$ fits the estimated monopole well. 

\cref{fig:ml_Nant<7>_Npoly<3>_achrom_idx<0>andNant<7>_Npoly<3>_achrom_idx<10>} (right) shows the inferred 21-cm signal when the foreground correction features 10\% errors. The foreground and 21-cm monopole GLS estimation errors are larger than the perfect foreground correction case, owing to the imperfectly modelled foreground modes. Correspondingly, the 21-cm signal features weaker constraints, with a $1\sigma$ width of $\sim 20$ mK at 70 MHz. This result is approximately equivalent to the achromatic single-spectrum fitting shown in \cref{fig:showcase_binwise} (top-left), which also uses a polynomial of order 3. Therefore, the performance of the mapmaking method for data taken achromatically approximately matches a simple SSF, when the foreground correction features $\sim 10\%$ errors.

\subsubsection{Chromatic case}
\label{sec:chrom ML results}
\begin{figure}
    \centering
    \includegraphics[width=\columnwidth]{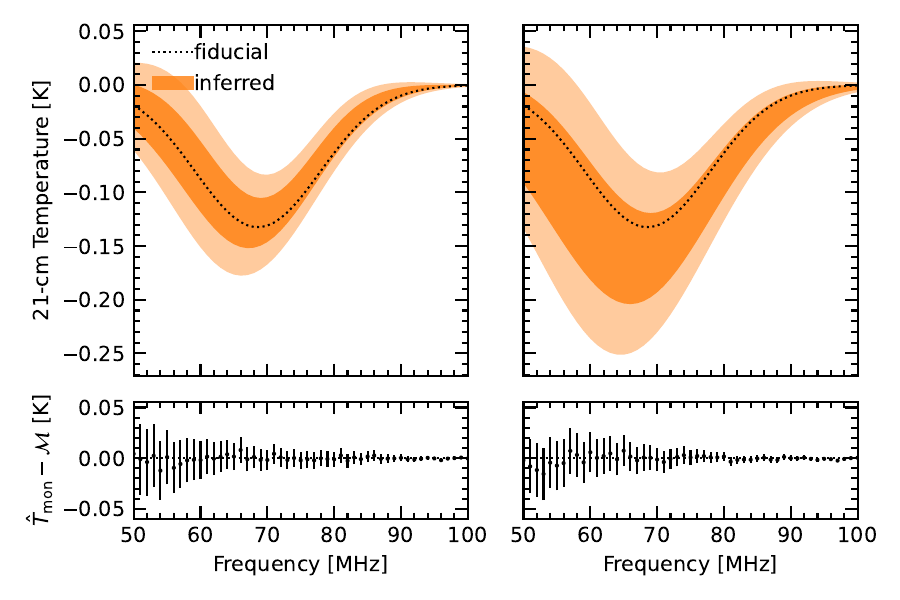}
    \caption{Results from the mapmaking method (c.f. \cref{fig:ml_Nant<7>_Npoly<3>_achrom_idx<0>andNant<7>_Npoly<3>_achrom_idx<10>}) for data collected with chromatic beams with $c=1.6\times10^{-2}$ (left) and $3.4\times10^{-2}$ (right), where unmodelled modes have been corrected imperfectly with $\sigma_T(70\,\mathrm{MHz})=10\%$. Both MCMC monopole fits have been carried out with $N_\mathrm{poly}=3$.}
    \label{fig:ml_Nant<7>_Npoly<3>_chrom<1.6e-02>_idx<10>andNant<7>_Npoly<3>_chrom<3.4e-02>_idx<10>}
\end{figure}

\cref{fig:ml_Nant<7>_Npoly<3>_chrom<1.6e-02>_idx<10>andNant<7>_Npoly<3>_chrom<3.4e-02>_idx<10>} shows the inferred 21-cm iso-probability contours for data taken with chromatic beams with curvature coefficients of $1.6\times10^{-2}$ (left) and $3.4\times10^{-2}$ (right), when the higher-mode foreground correction applied features $\sim10\%$ errors. In the lower-chromatic curvature case, the $1\sigma$ inference width of the 21-cm signal is approximately the same as the achromatic case with a  $\sim10\%$ error higher mode foreground correction. In the higher-curvature case, the inference width is approximately double this, while the error on the monopole estimate is similar. Both of these cases successfully infer the correct shape and magnitude of the 21-cm signal, unlike the corresponding chromaticity cases for the BFCC SSF shown in \cref{fig:showcase_binwise} (upper-right and lower-left).

\begin{figure}
    \centering
    \includegraphics[width=\columnwidth]{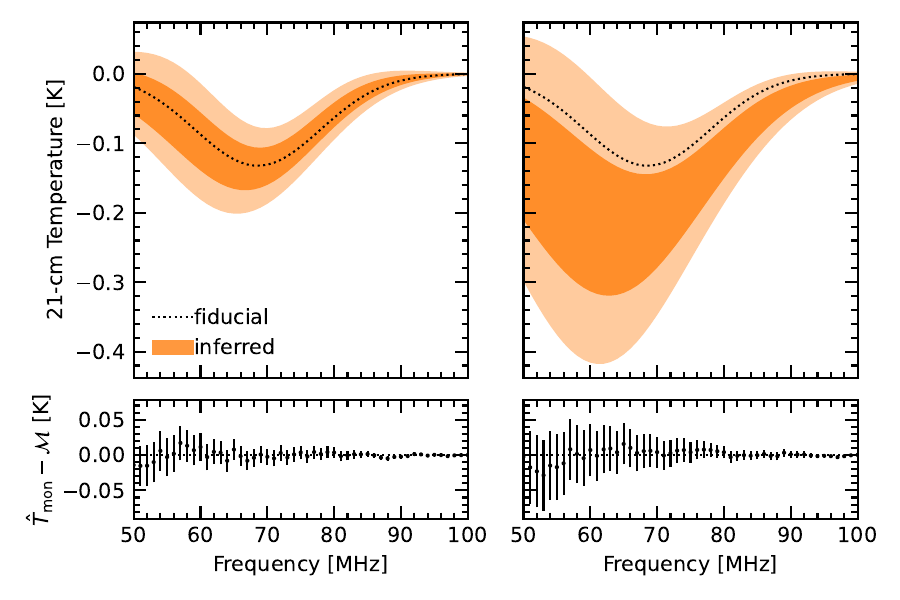}
    \caption{Results from the mapmaking method (c.f. \cref{fig:ml_Nant<7>_Npoly<3>_achrom_idx<0>andNant<7>_Npoly<3>_achrom_idx<10>}) for data collected with ``very chromatic'' beams with $c=5.2\times10^{-2}$  (left), and chromatic beams with $c=3.4\times10^{-2}$ (right), where unmodelled modes have been corrected imperfectly with $\sigma_T(70\,\mathrm{MHz})=10\%$ (left) and ``very imperfectly'' with $\sigma_T(70\,\mathrm{MHz})=20\%$ (right). MCMC monopole fits have been carried out with $N_\mathrm{poly}=3$ (left) and $N_\mathrm{poly}=4$ (right).} 
    \label{fig:ml_Nant<7>_Npoly<3>_chrom<5.2e-02>_idx<10>andNant<7>_Npoly<4>_chrom<3.4e-02>_idx<20>}
\end{figure}

We push the model further in \cref{fig:ml_Nant<7>_Npoly<3>_chrom<5.2e-02>_idx<10>andNant<7>_Npoly<4>_chrom<3.4e-02>_idx<20>} (left), by observing with a beam of an even higher chromaticity curvature, $c=5.2\times10^{-2}$, and \cref{fig:ml_Nant<7>_Npoly<3>_chrom<5.2e-02>_idx<10>andNant<7>_Npoly<4>_chrom<3.4e-02>_idx<20>} (right) using $c = 3.4\times 10^{-2}$ and a higher-mode foreground correction with 20\% errors. Increasing the chromaticity curvature further makes little difference to the inferred signal. However, we find that introducing higher foreground correction error both biases the inferred 21-cm signal and significantly broadens the iso-probability contours to well beyond the magnitude of the signal. The presence of the signal is still inferred to 2$\sigma$, however it would be difficult to conclude that such a result were not a feature of unknown errors.

\begin{figure}
    \centering
    \includegraphics[width=\columnwidth]{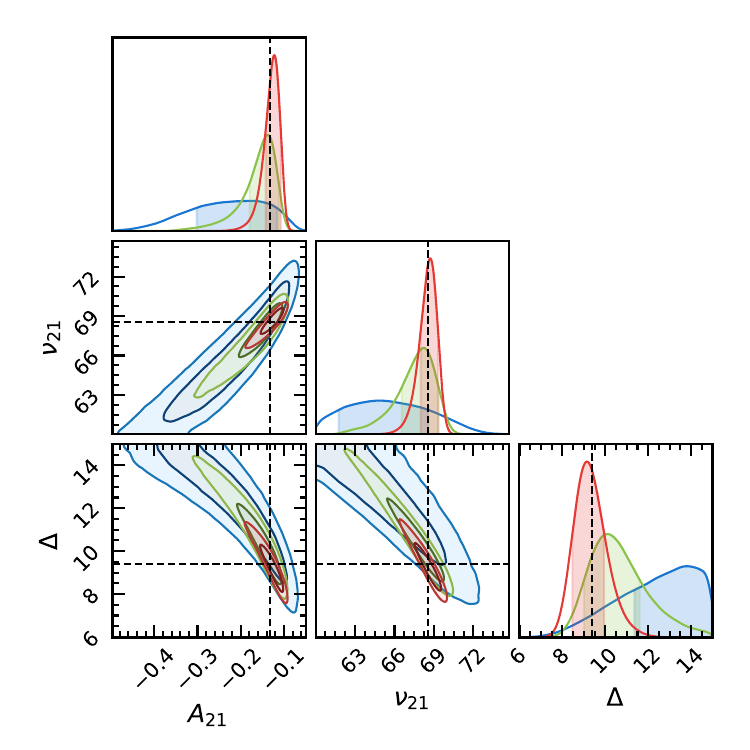}
    \caption{The marginalised 21-cm posteriors of the mapmaking method, corresponding to the three right-hand cases in \cref{fig:ml_Nant<7>_Npoly<3>_achrom_idx<0>andNant<7>_Npoly<3>_achrom_idx<10>,fig:ml_Nant<7>_Npoly<3>_chrom<1.6e-02>_idx<10>andNant<7>_Npoly<3>_chrom<3.4e-02>_idx<10>,fig:ml_Nant<7>_Npoly<3>_chrom<5.2e-02>_idx<10>andNant<7>_Npoly<4>_chrom<3.4e-02>_idx<20>}, i.e. the achromatic case with $\sigma_T(70\,\mathrm{MHz})=10\%$, the $c=3.2\times10^{-2}$ case with $\sigma_T(70\,\mathrm{MHz})=10\%$, and the $c=3.2\times10^{-2}$ case with $\sigma_T(70\,\mathrm{MHz})=20\%$. These cases respectively correspond to the smallest (red) contour, the middle-size (green) contour and the largest (red) contour.}
    \label{fig:showcase_ml_corner}
\end{figure}

The marginalised 21-cm posteriors for the achromatic case with 10\% foreground correction errors, the $c=3.4\times10^{-2}$ chromatic case with 10\% foreground correction errors and the $c=3.4\times10^{-2}$ chromatic case with 20\% foreground correction errors are shown in \cref{fig:showcase_ml_corner}. The posteriors show the same trend as the individual inference figures, with a combination of chromatic beams and poorer higher-mode foreground corrections yielding poorer constraints on the 21-cm parameters.
\section{Discussion}
\label{sec:discussion}
\subsection{Comparison to other methods}
\begin{table*}
\centering
\begin{tabular}{@{}lllll@{}}
\toprule
\textbf{Method} & \textbf{Publications} & \textbf{Required information} & \textbf{Reference map errors} & \textbf{Beam errors} \\ \midrule
\textbf{BFCC} & \cite{2017:MozdzenBowmanMonsalve,2019:MozdzenMaheshMonsalve} & $T_\mathrm{ref}(\nu;\mathbf{n})$, $f_\mathrm{beam}(\nu;\mathbf{n})$ & \multicolumn{1}{c}{N} & \multicolumn{1}{c}{N} \\
\textbf{BFCC + data model} & \cite{2024:PaganoSimsLiu} & $T_\mathrm{ref}(\nu;\mathbf{n})$, $f_\mathrm{beam}(\nu;\mathbf{n})$ & \multicolumn{1}{c}{Y} & \multicolumn{1}{c}{N} \\
\textbf{N-regions} & \cite{2021:AnsteydeLeraAcedoHandley,2023:AnsteydeLeraAcedoHandley,2024:PaganoSimsLiu} & $T_\mathrm{ref}(\nu_\mathrm{ref};\mathbf{n})$, $f_\mathrm{beam}(\nu;\mathbf{n})$ & \multicolumn{1}{c}{Y} & \multicolumn{1}{c}{N} \\
\textbf{\texttt{pylinex}} & \cite{2018:TauscherRapettiBurns,2020a:TauscherRapettiBurns} & $\{T_{\mathrm{ref},i}(\nu;\mathbf{n})$, $f_{\mathrm{beam},i}(\nu;\mathbf{n})\}$ & \multicolumn{1}{c}{Y} & \multicolumn{1}{c}{Y} \\
\textbf{mapmaking} & [this work] & $\boldsymbol\mu'', \mathbf{C}''_a$, $f_\mathrm{beam}(\nu;\mathbf{n})$ & \multicolumn{1}{c}{Y} & \multicolumn{1}{c}{N} \\ \bottomrule
\end{tabular}
\caption{Comparison of a number of 21-cm signal extraction methods, including the mapmaking method presented in this work. The ``Required information'' column lists the information each model uses in order to correct for beam-chromatic coupling effects. The rest of the columns denote the capabilities of each model to deal with different types of errors in the information provided.}
\label{table:21cm extraction methods}
\end{table*}

\begin{figure}
    \centering
    \includegraphics[width=\linewidth]{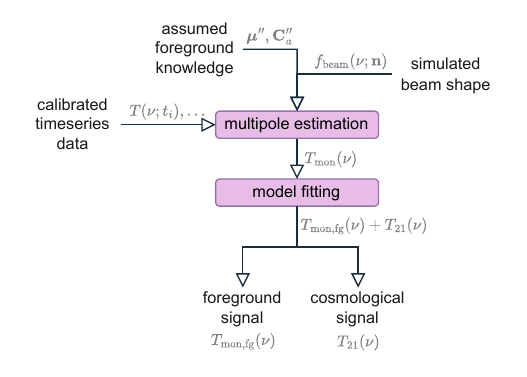}
    \caption{A flowchart showing the mapmaking 21-cm signal extraction method. Absolute-calibrated timeseries spectral data from a global array of 21-cm experiments is ran through a multipole estimation step, which recovers the monopole of the 21-cm signal and foregrounds. This step requires the mean ($\boldsymbol{\mu''}$) and covariance $\mathbf{C}_a''$ of a probability distribution describing the higher order spherical harmonic modes of the foregrounds. The resulting single spectrum is free of beam-chromatic foreground coupling, so it can be fit with the sum of a low-order polynomial foreground model and a 21-cm signal model to recover the cosmological signal.}
    \label{fig:mapmaking pipeline}
\end{figure}

The mapmaking method developed in this work is represented in the flowchart \cref{fig:mapmaking pipeline}. This is similar to the BFCC method, with the multipole estimation step in the mapmaking method playing the role of the chromaticity correction step in the BFCC method. We have shown that the mapmaking-based method succeeds in inferring the 21-cm global signal from data observed by chromatic antennas. In comparison, the standard BFCC SSF applied to the same data fails for any amount of chromaticity we tested. This performance gain over the BFCC method is in line with other methods developed in order to account for chromatic beam effects, namely the BFCC data model approach \cite{2023:SimsBowmanMahesh}, the N-regions model \cite{2021:AnsteydeLeraAcedoHandley} and {\small PYLINEX} \cite{2018:TauscherRapettiBurns}. 

These other methods are similar to the mapmaking method in that they require information about the beam and some reference foreground angular shape (in the case of all methods) and frequency shape (in the case of all but the N-regions method). The exact information required is summarised in \cref{table:21cm extraction methods}. When used to fit beam-chromatic data, the different methods are resilient to errors in the provided information to various degrees. The BFCC method requires a reference foreground map which exactly matches the fiducial foregrounds at all frequencies in the measured range. The BFCC data model relaxes this by requiring the reference foreground map to only match the fiducial foregrounds at a single frequency within the measured range. The N-regions model is even more relaxed, requiring a foreground map at a single frequency outside of the measured range which need not exactly match the fiducial foregrounds at that frequency. 

{\small PYLINEX} and the mapmaking method presented here differ to these methods, as they operate with the requirement of a distribution of reference foreground instances which must represent the fiducial foregrounds. In the case of {\small PYLINEX}, this consists of a training set of many foreground simulations in the measured frequency range. In the case of the mapmaking method, this consists of a probability distribution describing the $l\gtrsim 5$ spherical harmonic modes of the fiducial foregrounds. This may constitute an improvement over the {\small PYLINEX} method, as information describing the largest-magnitude modes need not be provided.

While the N-regions method is quite different to the method we present, they are similar in that they both infer a physical representation of the foreground sky. Both models rely on the prior knowledge of small angular scale foreground information, which the N-regions model uses by scaling it to lower frequencies in pixel space. Additionally, both models have been demonstrated to work when the foreground information provided features errors on the order of 10\%.

Currently, it is difficult to compare the foreground correction error in this work to results obtained with the {\small PYLINEX} method. {\small PYLINEX} could in principle be used with a stochastic foreground model with well-defined errors to form its beam-weighted foreground modelling set. However, \cite{2020a:TauscherRapettiBurns} observe the same 2BE foreground model with varyingly-chromatic beams, while \cite{2023:SaxenaMeerburgdeLeraAcedo} generate variation in the foregrounds by permuting spectral index values of a 2BE-like model, instead of injecting percentage errors. 

It is additionally important to note the capability of the {\small PYLINEX} method in dealing with errors in the provided beamfunction. This is made possible as the method relies on a set of instances of beam-weighted foregrounds. If the foreground instances are observed with varying beam instances, the method will be flexible to this variability. We leave the investigation of the effect of beam errors on the mapmaking method for further work.

\subsection{Foreground Model Gaussianity}
\label{sec:discussion_gauss}
The maximum-likelihood inversion we have performed in this paper implicitly assumes that the $l>l_\mathrm{mod}$  spherical harmonic modes of the true foregrounds may be well-described by a Gaussian distribution. The S2BE model we use in this work generates modes which are approximately Gaussian. In order to show this, we take the spherical transform of \cref{eq:approx_frac_err} and truncate it to higher-order modes by multiplying by $Y''^{-1}$, resulting in
\begin{equation}
\begin{aligned}
    &\sum_pY''^{-1}_{\alpha p} T^{(i)}_{\mathrm{S2BE},p} \approx \sum_pY''^{-1}_{\alpha p} T_{\mu,p} + \sum_pY''^{-1}_{\alpha p} T_{\mu,p} \delta_p \log \left(408/\nu_i\right) \,, \\[5pt]
    &\Delta''_\alpha \approx     \sum_pY''^{-1}_{\alpha p} T_{\mu,p} \delta_p \log \left(408/\nu_i\right)\,,
\end{aligned}
\end{equation}
i.e. each component of $\boldsymbol\Delta''$ is approximately a weighted sum of $\boldsymbol\delta$. Since these are all independent and Gaussian distributed, in the small-$\sigma_\delta$ limit, $\boldsymbol\Delta''$ is Gaussian distributed.

\cite{2024:PaganoSimsLiu} use a similar extension of the 2BE model in order to simulate foreground model uncertainty. The authors introduce Gaussian fractional errors to each pixel of the $\T_{408}$ basemap in \cref{eq:GSMA}, instead of deviating the power law indices. They smooth these error instances with a Gaussian beam in order to introduce spatial correlations. As \cref{eq:GSMA} linearly rescales the basemap, and both convolution and spherical harmonic decomposition preserve the Gaussianity of the fractional errors, $\boldsymbol\Delta''$ computed using their foreground model would be Gaussian-distributed.

A full Bayesian treatment of well-calibrated radio sky surveys such as that promised by the B-GSM may not exhibit Gaussian $l>l_\mathrm{mod}$  spherical harmonic modes. In this case, the minimization of the chi-squared function in \cref{eq:split obs chi-sq} would not yield the optimum estimate of the sky spherical harmonics. It is possible that the likelihood would still be analytically tractable. If this were not the case, methods such as variational inference with normalising flows \citep{2015:JimenezMohamed} could be explored, as a way of approximating a non-Gaussian likelihood featuring a large parameter space, as in this paper with $N_{l\mathrm{mod}}=36$.

\subsection{Beams}
\label{sec:discussion_beams}
We have used a simple beam model in this paper which features curvature in its chromatic profile. It does not feature the high frequency spectral oscillations common to real dipole antennas. However it has been shown previously \citep{2020a:TauscherRapettiBurns} that curvature in the beam chromatic profile is sufficient to significantly bias the typical BFCC SSF. The mapmaking method remains robust to this. Modifying the beam matrix to simulate more complex beam variations with frequency is straightforward, thus the formalism is readily applicable to more realistic chromaticity profiles.

We have additionally assumed that all antennas in the array have the same chromaticity profile. Even if the experiments were all identical, differing environmental conditions at the different sites would alter the beam chromaticity of each antenna. Different beam profiles may readily be represented in our formalism with slight modification. Currently, the sky in each frequency bin is convolved with the same beam. The pointing matrices corresponding to the pointings of each antenna are then vertically stacked to form $P$. This corresponds to an observation of the sky as if it were all taken by the same experiment with the same beam. Measurements of the sky with two antennas in different locations with different beams can be represented as two sets of $N_\mathrm{freq}$ observation equations of the form of \cref{eq:cmb_mother}, with different pointing matrices and beam matrices. These two sets of $N_\mathrm{freq}$ equations can be block-diagonalised into a single equation of the form \cref{eq:cmb_mother block} as before, with $2N_\mathrm{freq}$ blocks.

Finally, the beam profile in this work is assumed to be azimuthally symmetric. This allows the observation equation to consist of a beam-convolution of the sky, followed by matrices which pick out particular pixels in order to convert to timeseries data. It is possible to write an observation equation for a non-azimuthally symmetric beam in terms of the spherical harmonics of the beam and the observed sky \citep{2001:WandeltGorski}. The resulting observation equation would no longer be structured in the block-separated fashion of \cref{eq:cmb_mother block}, as the beam convolution of the sky would depend on the angle of the beam about the zenith. Despite this change, it is still possible to write the observation equation in the form $\d = \mathbf{A}\a + \n$, allowing the application of the same mapmaking methodology described in \cref{sec:GLS}.

\section{Conclusion}
\label{sec:conc}
Beam chromatic foreground coupling is a major difficulty faced in the calibration of 21-cm global experiments. In this paper, we have presented a novel application of CMB mapmaking techniques to 21-cm global signal extraction, developing a novel method which estimates the chromaticity-free monopole of the 21-cm sky before using a standard single-spectrum fit to separate the foregrounds and the signal. In this way, the data is pre-cleaned of chromatic effects and contributions from non-monopole foreground modes, allowing a simple SSF with few polynomial terms to adequately fit the foregrounds and signal.

The method we present provides a substantial improvement over the standard BFCC method in mitigating the effects of beam chromaticity. The BFCC method relies on the absolute accuracy of a foreground template. It can therefore fail to correct for beam-chromatic foreground coupling when the true foreground spatial or spectral structure does not align with the template. In contrast, the mapmaking method we presented relies on a representative probability distribution of the true foregrounds. We show that confident signal inference is possible given a fractional uncertainty in the foreground model on the order of 10\%. While a fully calibrated probabilistic model of the foregrounds based on radio sky surveys is not yet available, the B-GSM promises to make this requirement soon a reality.

In order to perform the monopole estimation step, we require data from a number of latitudinally-separated 21-cm antennas. Combining the data streams of separate antennas has previously been considered, with the REACH experiment having deployed a blade-dipole antenna in the Karoo Desert of South Africa ($\sim 30\degr$ S), with plans to deploy a conical log-spiral antenna in a similar location. The two antenna designs have been chosen to feature differing responses. When combined, the data from the two antennas has been shown to better constrain the 21-cm signal in simulations, assuming that the beamfunctions of both antennas are well-known \citep{2023:AnsteydeLeraAcedoHandley,2023:SaxenaMeerburgdeLeraAcedo}. 

There are also a number of 21-cm global experiments which feature multiple spatially-separated deployment sites around the globe, with an aim to better control environmental and foreground systematics. These include MIST, which does not feature a ground-plane, allowing for ease of transportation and deployment. MIST has been tested in California, Nevada (both $ 37\degr$ N) and the High Canadian Arctic ($79\degr$ N) \citep{2024:MonsalveAltamiranoBidula}. Additionally, the EDGES-3 experiment has been deployed in Inyarrimanha Illgari Bundara, the Murchison Radio-astronomy Observatory (MRO) site in Western Australia ($26\degr$ S) and Devon Island ($75\degr$ N) in the Canadian Arctic \citep{edges_memo_396}. Adak, Alaska ($\sim 51\degr$ N), is being considered as an upcoming deployment site \citep{edges_memo_470}. While all of these experiments taken together represent a large spread in observing latitudes, the possibility of combining the data streams of different instruments, which feature different calibration methods and systematics, remains an open question.

Our approach does not intrinsically require multiple antennas. The same approach could be applied to an experiment that actively scanned the sky, perhaps a horn experiment \citep[e.g.][]{2024arXiv241000076B}, so mounted as to be pointable rather than zenith tracking. It will be particularly interesting to explore the application of the mapmaking method to future satellite based experiments such as DSL \citep{2021:ChenYanDeng,2022:ShiDengXu}, DAPPER \citep{2021:Burns}, or PRATUSH \citep{2023ExA....56..741S}. A satellite scan pattern could allow data to be taken with full sky coverage by a single experiment, while avoiding spatially and temporally varying ionospheric contaminants which would otherwise risk biasing the inference of sky multipoles.

\appendix
\section{Means and Covariances of the S2BE Model}\label{apdx:mean_corr}
We consider probability density function (PDF) of a single pixel $p$ of the S2BE model minus the CMB temperature at the frequency $\nu_i$, which we denote $\Delta T_{\mathrm{S2BE},p} = T_{\mathrm{S2BE},p}^{(i)} -T_\mathrm{cmb}$. We also denote the corresponding pixel of the 2BE model minus the CMB temperature at frequency $\nu_i$ as $\Delta T_{\mathrm{2BE},p} = T_{\mathrm{2BE},p}^{(i)} -T_\mathrm{cmb}$. The PDF is given by the standard change of variables formula
\begin{equation}
\begin{aligned}
    f(\Delta T_{\mathrm{S2BE},p}) &= G\left(\delta_p;\mu=0, \sigma=\sigma_\delta\right) \left| \frac{\mathrm{d}\delta_p}{\mathrm{d}\Delta T_{\mathrm{S2BE},p}} \right| \\[5pt]
    &= \frac{1}{\Delta T_{\mathrm{S2BE},p}\sqrt{2\pi} \sigma_T} \exp \left\{ -\frac{1}{2\sigma_T^2}\left[ \log \left(\frac{\Delta T_{\mathrm{S2BE},p}}{T_{408,p}-T_\mathrm{cmb}}\right) \right.\right. \\
   &\qquad \left.\left. - \gamma_p \log \left(\frac{408}{\nu_i}\right) \right]^2 \right\}\,,
   \label{eq:prob_dist}
\end{aligned}
\end{equation}
where $G$ is the mean-zero normal distribution from which $\delta_i$ is drawn. The pixel mean of the model is computed by taking the expectation value of the distribution
\begin{equation}
\begin{aligned}
    \langle \Delta T_{\mathrm{S2BE},p}\rangle &= T_{\mu,p}^{(i)} + T_\mathrm{cmb}= \int_0^\infty \Delta T_{\mathrm{S2BE},p} f(\Delta T_{\mathrm{S2BE},p}) \, \mathrm{d}(\Delta T_{\mathrm{S2BE},p}) \\
    &= \left( T_{\mathrm{2BE},p} - T_\mathrm{cmb}  \right) \exp\left(\sigma_T^2/2\right) \\[5pt]
    \Rightarrow T_{\mu,p}^{(i)} &= \left( T_{\mathrm{2BE},p} - T_\mathrm{cmb}  \right) \exp\left(\sigma_T^2/2\right) + T_\mathrm{cmb} \,.
    \label{eq:S2BE mean calc}
\end{aligned}
\end{equation}
Similarly, the expectation of the square is 
\begin{equation}
\begin{aligned}
    \langle \Delta T_{\mathrm{S2BE},p}^2\rangle &= \int_0^\infty \Delta T_{\mathrm{S2BE},p}^2 f(\Delta T_{\mathrm{S2BE},p}) \, \mathrm{d}(\Delta T_{\mathrm{S2BE},p}) \\
    &= \left(T_{\mathrm{2BE},p} - T_\mathrm{cmb}  \right)^2 \exp\left(2\sigma_T^2\right)\,.
     \label{eq:S2BE meansq calc}
\end{aligned}
\end{equation}

The covariance matrix of $\Delta \T_{\mathrm{S2BE}}$ is the same as the covariance matrix of $\T_\mathrm{S2BE}$, since they only differ by a constant offset. Additionally, the covariance matrix will be diagonal as all pixels are uncorrelated. The diagonal entries of the covariance matrix for $\T_\mathrm{S2BE}$ are therefore given by
\begin{equation}
    C_{T,pp}^{(i)} = \langle \Delta T_{\mathrm{S2BE},p}^2\rangle -  \langle \Delta T_{\mathrm{S2BE},p}\rangle^2 \,,
\end{equation}
which, substituting \cref{eq:S2BE mean calc,eq:S2BE meansq calc}, evaluates to \cref{eq:temp covar}. This covariance matrix may be transformed to represent the covariance of the spherical harmonic representation of the model
\begin{equation}
\begin{aligned}
    \mathbf{C}_{a} &\equiv \langle \a_\mathrm{S2BE} \a_\mathrm{S2BE}^T\rangle - \langle \a_\mathrm{S2BE}\rangle \langle \a_\mathrm{S2BE} \rangle^T \\
    &= \mathbf{Y}^{-1} \left(\langle \T_\mathrm{S2BE} \T_\mathrm{S2BE}^T\rangle - \langle \T_\mathrm{S2BE}\rangle \langle \T_\mathrm{S2BE} \rangle^T \right) (\mathbf{Y}^{-1})^T \\
    &= \mathbf{Y}^{-1} \mathbf{C}_T (\mathbf{Y}^{-1})^T \,,
\end{aligned}
\end{equation}
as $\a_\mathrm{S2BE}\equiv\mathbf{Y}^{-1}\T_\mathrm{S2BE}$. If the relation $\a_\mathrm{S2BE}''\equiv\mathbf{Y}''^{-1}\T_\mathrm{S2BE}$ is used instead, this results in the covariance matrix of the corrected modes, $\mathbf{C}_{a}''$. Since $\boldsymbol\Delta''$ and $\a_\mathrm{S2BE}''$ differ by a constant offset, this is also the covariance matrix of $\boldsymbol\Delta''$.

\section*{Acknowledgements}
YI acknowledges the support of a Science and Technology Facilities Council studentship [grant number ST/W507519/1]. JRP acknowledges support from STFC grant ST/Y004132/1. Some of the results in this paper have been derived using the {\small HEALPY} and {\small HEALPIX} package. We thank Boris Leidstedt and Alan Heavens for useful discussions.

\section*{Data Availability}
The data underlying this article will be shared on reasonable request to the corresponding author. The implementation of the mapmaking method, as well as the code used to generate the figures in this paper are publicly available at \url{https://github.com/YordanIg/globmaps}.


\bibliographystyle{mnras}
\bibliography{bibliography}

\bsp	
\label{lastpage}
\end{document}